\documentclass[reprint,amsmath,amssymb,aps]{revtex4-1}

\usepackage{graphicx}
\usepackage{dcolumn}
\usepackage{bm}
\usepackage[colorlinks=true,linkcolor=red,urlcolor=blue,citecolor=blue]{hyperref}

\usepackage{multirow}
\usepackage{subfigure}
\usepackage{gensymb}
\usepackage{acronym}
\acrodef{GW}{gravitational wave}
\acrodef{DWD}{white dwarf binarie}
\acrodef{WD}{white dwarf}
\acrodef{GR}{general relativity}
\begin{document}

\title{Sky location of Massive Black Hole Binaries in the foreground of Galactic white dwarf binaries}

\author{Pan Guo$^{1,4}$}
\author{Hong-Bo Jin$^{1,2,3}$}\email[]{Corresponding author Email: hbjin@bao.ac.cn}
\author{Cong-Feng Qiao$^{1,4}$}
\author{Yue-Liang Wu$^{1,4,5}$}
\affiliation{\begin{footnotesize}	
		${}^1$The International Centre for Theoretical Physics Asia-Pacific, University of Chinese Academy of Sciences, Beijing 100190, China \\
		${}^2$Institute for Frontiers in Astronomy and Astrophysics, Beijing Normal University, Beijing, China\\
		${}^3$National Astronomical Observatories, Chinese Academy of Sciences, Beijing 100101, China \\
		${}^4$The School of Physical Sciences, University of Chinese Academy of Sciences, Beijing 100049, China\\
		${}^5$Institute of Theoretical Physics, Chinese Academy of Sciences, Beijing 100190, China 
\end{footnotesize}}

\date{\today}
\begin{abstract}
For space-based gravitational wave (GW) detection, the main noise source for massive black hole binaries (MBHBs) is attributed to approximately $10^7$ double white dwarf binaries in the foreground.   
For a GW source, the amplitude of the detector response, recorded by a space-based gravitational wave detector, exhibits a modulation effect with a year period when observing the source from various orbital positions.
Under the adverse conditions mentioned above, where there is a strong foreground noise and annual modulation in the signals, we employed the wavelet transform and the strong-amplitude relevant orbital position search methods, which allows the weak MBHB sources to achieve higher locating accuracy. 
In detail, for two MBHB sources of lower intensity, the precision of luminosity distance, represented by the ratio $\Delta D_L / D_L$ at the 95$\%$ confidence level, is enhanced by factors of $\sim$ 2. And the angular resolutions, denoted by $\Delta \Omega_s$, are enhanced by a factor of $\sim$ 20.
These improvements increase the number of detectable GW sources, facilitates multi-messenger follow-up observations and provides constraints on the cosmological constant.

\end{abstract}
\maketitle

\section{Introduction}
\label{introduction}
During the O1-O3 observing runs, ground-based observatories from the LIGO, Virgo, and KAGRA collaborations \citep{KAGRA:2021vkt,KAGRA:2023pio} have detected nearly one hundred gravitational wave events with frequencies above 10 Hz. Gravitational wave detection from space-based observatories, such as LISA \citep{LISA:2017pwj}, Taiji \citep{Hu:2017mde}, and TianQin \cite{TianQin:2015yph,TianQin:2020hid}, operates in the low-frequency range, spanning from $10^{-4}$ to $10^{-1}$ Hz.
Gravitational wave (GW) signals emitted by the merger of massive black holes persist for an extended duration within the sensitive frequency band of space-based detectors, often overlapping with the continuous, nearly monochromatic GW signals produced by Galactic white dwarf binaries. The number of Galactic white dwarf binaries is on the order of $10^{7}$, and this type of source is also referred to as the foreground signal \citep{Farmer:2003pa, Nelemans:2009hy}. The number of resolvable foreground signals is about $10^{4}$ \citep{Littenberg:2020bxy,Zhang:2021htc}. 

The identification of foreground signals remains a challenge in the detection of space-based gravitational waves. Overlapping GW sources can interfere with the detection of other signals, including those from merging massive black hole binaries \citep{Klein:2015hvg}. There were many methods that have been proposed to isolate the DWDs \citep{Normandin:2018lzk} and \citep{Littenberg:2020bxy}.  The data analysis of those overlapping signals is one of the focuses of the LISA Data Challenges \citep{LDC_web} and the Taiji Data Challenge \citep{TDC_web}. The data analysis of overlapped GW signals has already become a general issue that needs to be addressed.

Compared with the double White Dwarf sources, massive black hole binaries (MBHBs) have higher intensity of response data on the detector. Moreover, signals from DWDs and MBHBs may also exhibit spectral overlap within the frequency spectrum. The resolvable foreground signals are the sources with relatively large amplitude and intensity, which will definitely affect the search of MBHB sources.

Sky localization of a gravitational source is a key scientific goal for gravitational wave observations\citep{Blaut:2011zz}. 
Quickly locating identified binaries is a basic requirement for space-based gravitational wave detection, which is used to calibrate detection data and reduce detector noise. 
Accurately determining the sky positions of these gravitational sources facilitates multi-messenger observations when combined with data from other gravitational wave detectors and astronomical observations. 
Additionally, the uncertainty in the sky localization parameter, specifically the luminosity distance, derived from these sources can contribute to refining the estimation of the Hubble constant in cosmological models through the distance-redshift relationship\cite{Song:2022siz}.
On cosmological parameters, the constraints from Taiji are similar with those from LISA and always tighter than those from TianQin\cite{Zhao:2019gyk}.
The LISA-Taiji network could significantly improve the constraint accuracies of cosmological parameters compared with the single Taiji mission\cite{Wang:2021srv,Wang:2020dkc}.
The Taiji-TianQin-LISA network alone could achieve a constraint precision of 0.9$\%$ for the Hubble constant, meeting the standard of precision cosmology\cite{Jin:2023sfc}.
These results imply that the accuracy of the sky location from a single detector is strongly sensitive to the constraint accuracies of cosmological parameters. Therefore, in this paper, based on single Taiji mission, the updated accuracies of the sky location of GW sources from the Q3-d model take into account the improvement of the accuracies of cosmological parameters.

For the sky location methods of gravitational wave signals, there are primarily two types: One method involves using the Fisher Information Matrix (FIM) to estimate the error in source parameters, providing an estimation of the spatial location of the gravitational source, particularly when there is a high signal-to-noise ratio (SNR). The inverse of the FIM yields the covariance matrix for these parameters.
Additionally, the angular resolution of the detector and the estimated signal parameters can be determined as functions of the source's spatial position and frequency.\citep{Wen:2010cr}. 
Jointly operating detectors can effectively achieve better positioning accuracy in the detection of gravitational waves\citep{Ruan:2019tje,Ruan:2020smc,Wang:2020vkg,Zhang:2020hyx}. The FIM method for parameter error estimation is limited by the SNR and provides only a general estimate of the detector's accuracy in locating gravitational sources.

The second kind of method is based on the Bayesian approach, where the posterior distribution describes the uncertainty of the gravitational source parameters. The most successful of these Bayesian approaches is based on Markov Chain Monte Carlo (MCMC) methods \citep{Crowder:2006eu,Crowder:2007ft,Littenberg:2011zg}. This method provides richer information about the source parameters, but compared with the first method, the computational cost is higher. However, the MCMC method is a classical algorithm for parameter estimation of gravitational sources and can perform complete parameter estimation for low-frequency GWs.


In our preceding research, as detailed in Ref. \cite{Guo:2023lzb}, the Taiji response function was formulated within the transverse-traceless gauge and subjected to numerical computation in the time domain. It is known that the amplitude of GW signals is modulated by the periodic motion of the GW detectors as the Earth orbits the Sun. Consequently, the detection of GW signals and the subsequent parameter estimation for GW sources must take into account the orbital positions of the detectors relative to the GW sources when analyzing data from space-based observatories. For verification binaries (VBs) signals, observation at the best orbital position of the detectors can improve the accuracy of sky location. In the paper\citep{Lin:2022huh}, the annual modulation produced by the galactic DWDs can provide useful information to distinguish the galactic foreground from stochastic gravitational wave backgrounds. 

When a MBHB merges, the signal amplitude is at its maximum. However, if the propagation direction of the gravitational waves is not perpendicular to the detector's arm, the signal's maximum amplitude may not be fully sensitive to the detectors. During the sky location search for MBHBs, the range of location parameters is constrained by the detectors' orbital positions. Consequently, when an MBHB with a sky location perpendicular to the detector's arm is merging, its signal is most strongly detected. We propose methods for searching the orbital positions that are relevant to strong amplitudes, which can help to reduce the search range for location parameters. At the optimal observational orbital position, the detector is best positioned to capture the merger signal from the MBHB wave source. Simultaneously, the detector arm's response signal intensity to the wave source reaches its peak.

Time-Delay Interferometry (TDI) is essential for missions similar to LISA to mitigate laser frequency noise. The response of a TDI combination to a gravitational wave signal is constructed by amalgamating the measurements obtained from time shifted laser links. The configuration of TDI-X of the first-generation TDI combinations\citep{armstrong1999time} is selected in this paper. 

In this paper, we focus on the sky location of MBHBs in the foreground of Galactic white dwarf binaries, considering the amplitude modulation effect of different detector positions on the response. The best short-time observations of MBHBs are presented in this paper. Three models for the populations of MBHBs, namely pop III, Q3-nod, and Q3-d \citep{Klein:2015hvg}, are considered to predict the events of massive black hole binary mergers. For some specific Q3-d sources,   the best sky location determination is achieved at the optimal detector position in this paper. 
To estimate the difference in location parameters, two scenarios are considered: one excluding and one including the DWDs foreground. The first scenario assumes an ideal case where we can remove 100$\%$ of the 10,000 loud DWD foreground signals. The second scenario represents the worst case, where none of these DWDs are removed.
For relatively strong Q3-d sources, the accuracy of the parameters obtained from overlapped signals is close to that obtained without overlapped signals. However, for MBHB sources of lower intensity, the accuracy of the parameters obtained from overlapped signals is much worse than that obtained without overlapped signals. The method of wavelet decomposition and reconstruction can significantly reduce the impact of overlap from DWDs for MBHB sources of lower intensity.  

The structure of this paper is as follows: in Sec.\ref{Sec:SM} , we mainly introduce the signal model of the MBHB and the DWDs sources; the detector response and the method used in this paper are shown in Sec.\ref{Sec:best_OB_Taiji}, the detector response is presented and best orbit position for observation is obtained 
; in Sec.\ref{Sec:SL_MBHB}, Sky location of MBHB at the best detector position is achieved; in Sec.\ref{Sec:conclusion} , there are the summary and conclusion of this paper.

\section{Signal Model}\label{Sec:SM}

\subsection{Massive Black Hole Binary}

Three models for the population of MBHB, i.e., pop III, Q3-nod, and Q3-d\citep{Klein:2015hvg}, are considered to predict the events of MBHB mergers. 
The parameters are obtained from the GW Toolbox, whose details are found at the paper\citep{Yi:2021wqf}. 
The detector parameters are set as follows, see the table ~\ref{tab:Detectors_parameters} for details.
\begin{table*}[!htbp]
    \centering
    \footnotesize
    \setlength{\tabcolsep}{1.6pt}
    \renewcommand{\arraystretch}{1.2}
    \begin{tabular}{lccc}
        \hline
         parameter & value & parameter & value  \\
        \hline
        the Arm length[m] & 3.0e9 & the Laser pow[W] & 2.2 \\
        Telescope diameter[m] & 0.35 & Accelaration Noise[Hz$^{-1}$] & 3.9e-44 \\
        Laser Shot Noise factor[Hz$^{-1}$] & 5.3e-38 & Other Optical Metrology system noise [Hz$^{-1}$] & 2.81e-38 \\
         The Time [days] & 365 & SNR-threshol & 8 \\
         Cosmology & Flat$\Lambda$CDM-Planck15 &  & \\ 
        \hline
    \end{tabular}
    \caption{Parameter list of Detectors.}
    \label{tab:Detectors_parameters}
\end{table*}

In this paper, the model Q3-d is selected among the three models, and parameters of 5 MBHB sources data are listed according to one year period. 
There are eight parameters:
intrinsic mass of primary $/$ secondary BH (M$_\odot$) $M1/M2$, cosmological redshift $z$, spin of primary$/$secondary BH	$s1/2$, luminosity distance (Gpc) $D_L$,	Ecliptic latitude (rad) $\beta$, Ecliptical longitude (rad) $\lambda$, and signal to noise ratio SNR. In this paper, the deflection Angle $\iota$ is set to 60 degrees, and the polarization Angle $\psi$ is set to 0.
The parameters are list In Table ~\ref{tab:Q3d_parameters}.
We select a higher-multipole gravitational waveform model for an eccentric binary black holes based on the effective-one-body-numerical-relativity formalism, waveform model SEOBNRE \citep{Cao:2017ndf,Liu:2021pkr}.
\begin{table*}[!htbp]
    \centering
    \footnotesize
    \setlength{\tabcolsep}{8pt}
    \renewcommand{\arraystretch}{1.5}
    \begin{tabular}{lccccccc}
        \hline
        Q3-d & $M1$ [M$_\odot$] & $M2$ [M$_\odot$] & $z$ & $D_L$ [Gpc] &  $\beta$ [rad] & $\lambda$ [rad]  & SNR \\
        \hline
        Q3-delay-1 & $209156$ & $27895$ & $7.25$ & $73.581$ & $-0.06470$ & $3.180$ & $82.54$ \\
        Q3-delay-2 & $1361322$ & $770150$ & $1.12$ & $7.870$ & $0.2066$ & $4.536$ & $1818$ \\
        Q3-delay-3 & $1272097$ & $1127338$ & $1.69$ & $12.986$ & $-0.7767$ & $3.965$ & $3053$ \\
        Q3-delay-4 & $389952$ & $103676$ & $5.54$ & $53.903$ & $0.9602$ & $5.167$ & $180.5$ \\
        Q3-delay-5 & $103676$ & $275282$ & $0.749$ & $4.758$ & $0.1067$ & $4.469$ & $3661$ \\
        \hline
    \end{tabular}
    \caption{Parameter list of model Q3-d.}
    \label{tab:Q3d_parameters}
\end{table*}

\subsection{Galactic white dwarf binaries}
For Taiji, since the frequency band for space-based GW detection ranges from $10^{-4}$ to $10^{-1}$ Hz, a large number of Galactic DWDs overlap in this range, forming what is known as the Galactic foreground.  
DWDs are described by a set of eight parameters: frequency $f$, frequency derivative  $\dot{f}$, amplitude ${\cal A}$, sky position in ecliptic coordinates $(\lambda,\beta)$, orbital inclination $\iota$, polarisation angle $\psi$, and initial phase $\phi_0$ \citep{Cutler:1997ta,Roebber:2020hso,Karnesis:2021tsh}. 
GWs emitted by a monochromatic source are characterized and quantified using the quadrupole approximation\cite{Landau:1962,Peters:1963}.
In this approximation, the GW signal is described  as a combination of the two polarizations ($+$ and $\times$): 
\begin{equation}\label{eq:hplus}
    h_+(t) = \mathcal{A}(1+\cos\iota^2)\cos( \Phi(t))
\end{equation}
\begin{equation}\label{eq:hcross}
    h_{\times}(t) = 2\mathcal{A}\cos\iota\sin(\Phi(t))
\end{equation}
with
\begin{equation}
    \Phi(t)=2\pi ft + \pi\dot{f}t^2 + \phi_0 
\end{equation}
\begin{equation}\label{eq:fdot}
    \dot{f}=\frac{96}{5}  \pi^{8/3} \left( \frac{G{\cal M}}{c^3} \right)^{5/3} f^{11/3},
\end{equation}
\begin{equation} \label{eq:GWamp}
    \mathcal{A} =\frac{2(G\mathcal{M})^{5/3}}{c^{4}D_L}(\pi f)^{2/3}, 
\end{equation}
where $\mathcal{M}\equiv (m_1 m_2)^{3/5}/(m_1+m_2)^{1/5}$ is the chirp mass, and $D_L$ is luminosity distance, and $G$ and $c$ are the gravitational constant and the speed of light, respectively.
The parameters of DWDs sources are obtained from the parameter distribution, which is detailed in the paper\cite{Guo:2023lzb}. 

\begin{figure*}
	\centering 
    \subfigure[Time domain]{
    \includegraphics[width=0.45\textwidth]{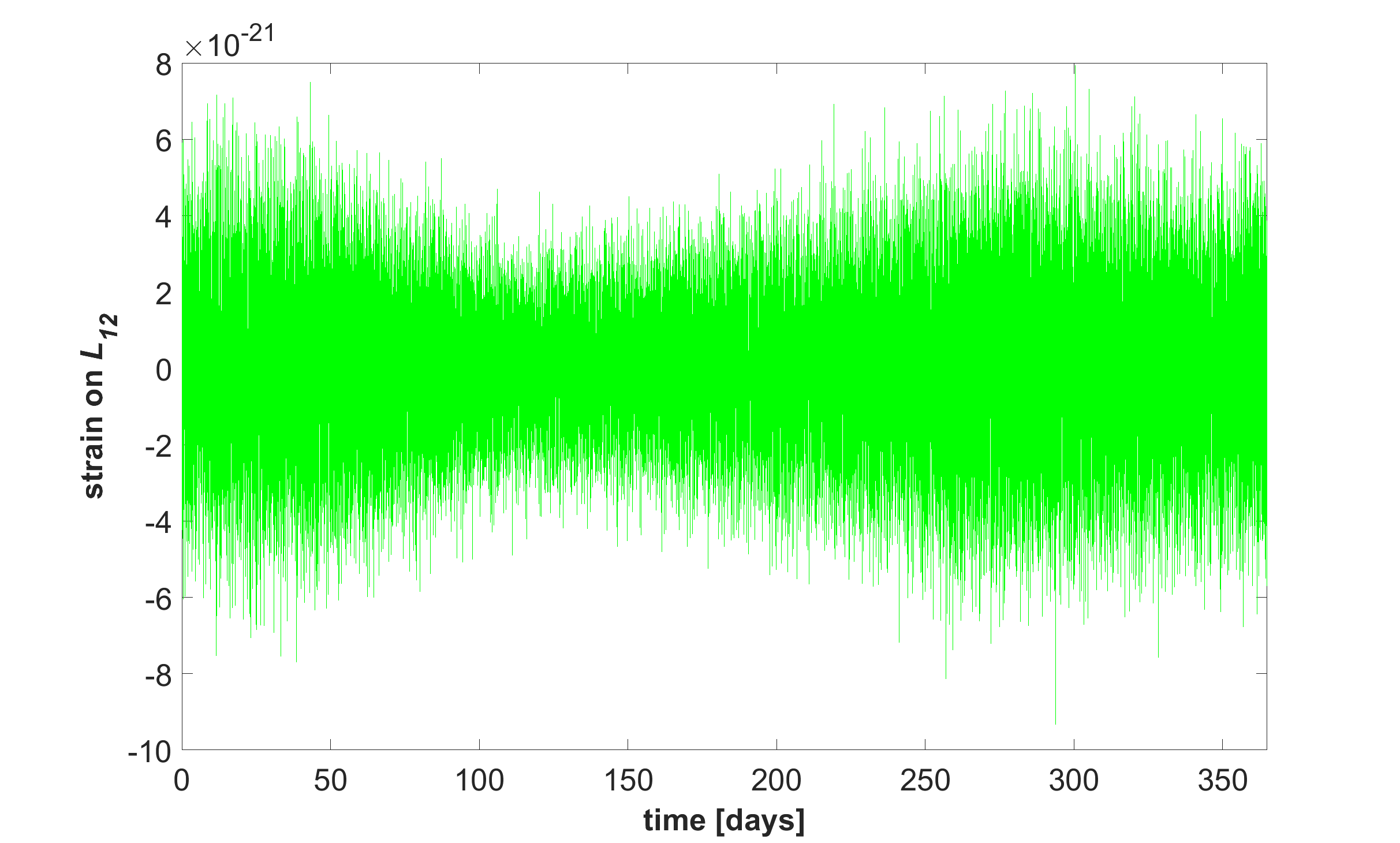}
    }
    \subfigure[Frequency domain]{\includegraphics[width=0.45\textwidth]{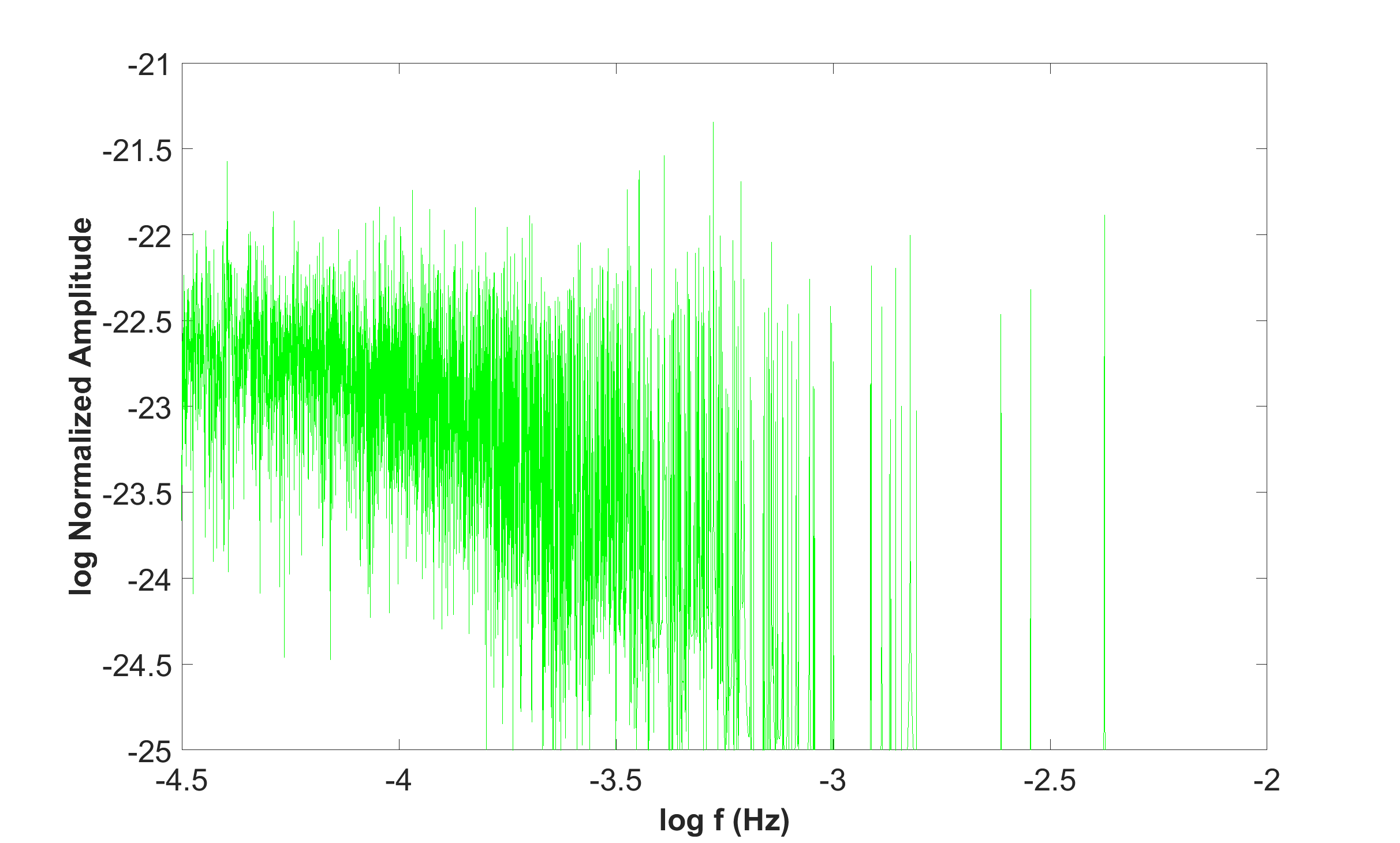}
    }
    \caption{The projection results on the detector $L_{12}$ of DWDs confusion.}
     \label{fig: DWDs_confusion_h12}
\end{figure*}

The coordinates under the $SSB$-frame are also composed of the spatial position coordinates ($\beta,\lambda$).
So we calculate the amplitude of each DWDs signal source at each the spatial position coordinates ($\beta,\lambda$) from the DWDs in the Synthetic Galactic Population. The formula for calculating the amplitude is the Eq.~ \ref{eq:GWamp}.
The latitude ($\beta\in \left[ -\pi/2,\pi/2\right]$) is divided into $100$ equal parts and the longitude ($\lambda \in \left[ 0,2\pi\right]$) is divided into $100$ equal parts. So we can get the grids of  $100\times 100$ on the sky map.
Because there are so many double white dwarf (DWD) binaries, there are numerous DWD signals within each spatial grid. Here, we select only the maximum value and retain its corresponding DWD binary as the source at that spatial location. Consequently, we obtain ten thousand DWD binaries with larger amplitudes. 
So we obtain ten thousand DWDs sources with larger amplitudes. 
Ten thousand DWDs sources are projected onto the detector arms ($L_{12}$, $L_{23}$, and $L_{31}$), the observation time is set as one year, and the sampling frequency is set to 0.1 Hz. The polarisation angle is set $\psi=0$, and initial orbital phase is set $\phi_0=0$.
Finally, all the projection signals are added together to get the DWDs sources confusion signal. The time and frequency domains of the DWDs sources confusion signal on the detector arm $L_{12}$ are as seen in Fig.~\ref{fig: DWDs_confusion_h12}. 

\subsection{Gravitational Wave Model}

This is a source frame, but for detectors, a Solar System Barycenter (SSB) frame, based
on the ecliptic plane is selected\footnote{\href{https://lisa-ldc.lal.in2p3.fr/static/data/pdf/LDC-manual-002.pdf}{https://lisa-ldc.lal.in2p3.fr/static/data/pdf/LDC-manual-002.pdf}}.In the $SSB$-frame, the standard spherical coordinates $\left(\theta,\phi\right)$, and the associated spherical orthonormal basis vectors $\left(\bf{e}r,\bf{e}\theta,\bf{e}\phi\right)$ are confirmed. The position of the
source in the sky will be parametrized by the ecliptice latitude $\beta =\pi/2 - \theta$ and the ecliptic 
longitude $\lambda = \phi$. The GW propagation vector $\hat{k}$ in spherical coordinates is now
\begin{equation}
    \hat{k} = -\bf{e} r = -\cos\beta\cos\lambda\,\hat{x} - \cos\beta\sin\lambda\,\hat{y} -
    \sin\beta\,\hat{z} \,.
\end{equation}
Introduce reference polarization vectors as
\begin{eqnarray}
    \hat{u} &=& -\bf{e}\theta = -\sin\beta\cos\lambda\,\hat{x} -\sin\beta\sin\lambda\,\hat{y} + \cos\beta\,\hat{z} \nonumber\\
    \hat{v} &=& -\bf{e}\phi = \sin\lambda\,\hat{x} - \cos\lambda\,\hat{y} 
\end{eqnarray}
The last degree of freedom between the frames corresponds to a rotation around the line of
sight, and is represented by the polarization angle $\psi$.
The polarization tensors are given by
\begin{eqnarray}
    {\mbox{\boldmath$\epsilon$}}^+ &=& \cos(2\psi) {\bf e}^+ - \sin(2\psi)
      {\bf e}^\times \nonumber\\
    {\mbox{\boldmath$\epsilon$}}^\times &=& \sin(2\psi) {\bf e}^+ +
    \cos(2\psi) {\bf e}^\times \, ,
\end{eqnarray}
where the
basis tensors ${\bf e}^+$ and ${\bf e}^\times$ are expressed in terms of two orthogonal unit vectors,
\begin{eqnarray}
    {\bf e}^+ &=& \hat{u} \otimes \hat{u} - \hat{v} \otimes \hat{v}
    \nonumber\\
    {\bf e}^\times &=& \hat{u} \otimes \hat{v} + \hat{v} \otimes \hat{u} \,.
\end{eqnarray}

GW traveling in the $\hat{k}$ direction are written as the linear combination of two independent polarization states,
\begin{equation}\label{wave}
    {\bf h}(\xi) = h_+(\xi) {\mbox{\boldmath$\epsilon$}}^+
     + h_\times(\xi) {\mbox{\boldmath$\epsilon$}}^\times \,,
\end{equation}
where the wave variable $\xi = t - \hat{k} \cdot {\bf x}$ gives the surfaces of constant phase. 

\section{Short-time Observation from Taiji}\label{Sec:best_OB_Taiji}

\subsection{Detector Response}

Like LISA, the individual Taiji spacecraft will follow independent Keplerian orbits\cite{Rubbo:2003ap}. The spacecraft positions are expressed as a function of time.  The parameter $\kappa$ gives the initial ecliptic longitude.    
 By setting the mean arm-length equal to those of the Taiji baseline, $L = 3 \times 10^9$ m.

The laser link signal is emitted from the spacecraft $i$ at $t_i$ moment and received by the spacecraft $j$ at $t_j$ moment. Assume that $c=1$. The gravitational wave travels  in  the $ \hat k$ direction and ${\mathbf x (t)}$ is the position on the laser link at time t.
The Eq.~(\ref{deltaL2}) can calculate the $\delta \ell_{i j}$ in a numerical way in the time domain\cite{Guo:2023lzb}.
\begin{equation} \label{deltaL2}
    \delta \ell_{i j} =\frac{1}{2} \hat{r}_{i j}(t) \otimes \hat{r}_{i j}(t) : \int_{t_i}^{t_j} {\bf h}(t-\hat{k} \cdot {\mathbf x (t)}) d t
\end{equation}
where,
$\hat{r}_{i j}(t)$ denotes the unit vector
\begin{equation}\label{rij}
    \hat{r}_{i j}(t_i) = \frac{{\bf x}_j(t_j) - {\bf
        x}_i(t_i)}{\ell_{i j}} \, .
\end{equation}
and ${\bf h}(\xi)$ is the gravitational wave tensor in the transverse-traceless gauge. 
The colon here denotes a double contraction, ${\bf a}:{\bf b} = a^{i j}b_{i j}$.

\subsection{Best orbit position for observation}
\begin{figure*}[!htbp]
	\centering 
    \subfigure[]{\includegraphics[width=0.3\textwidth]{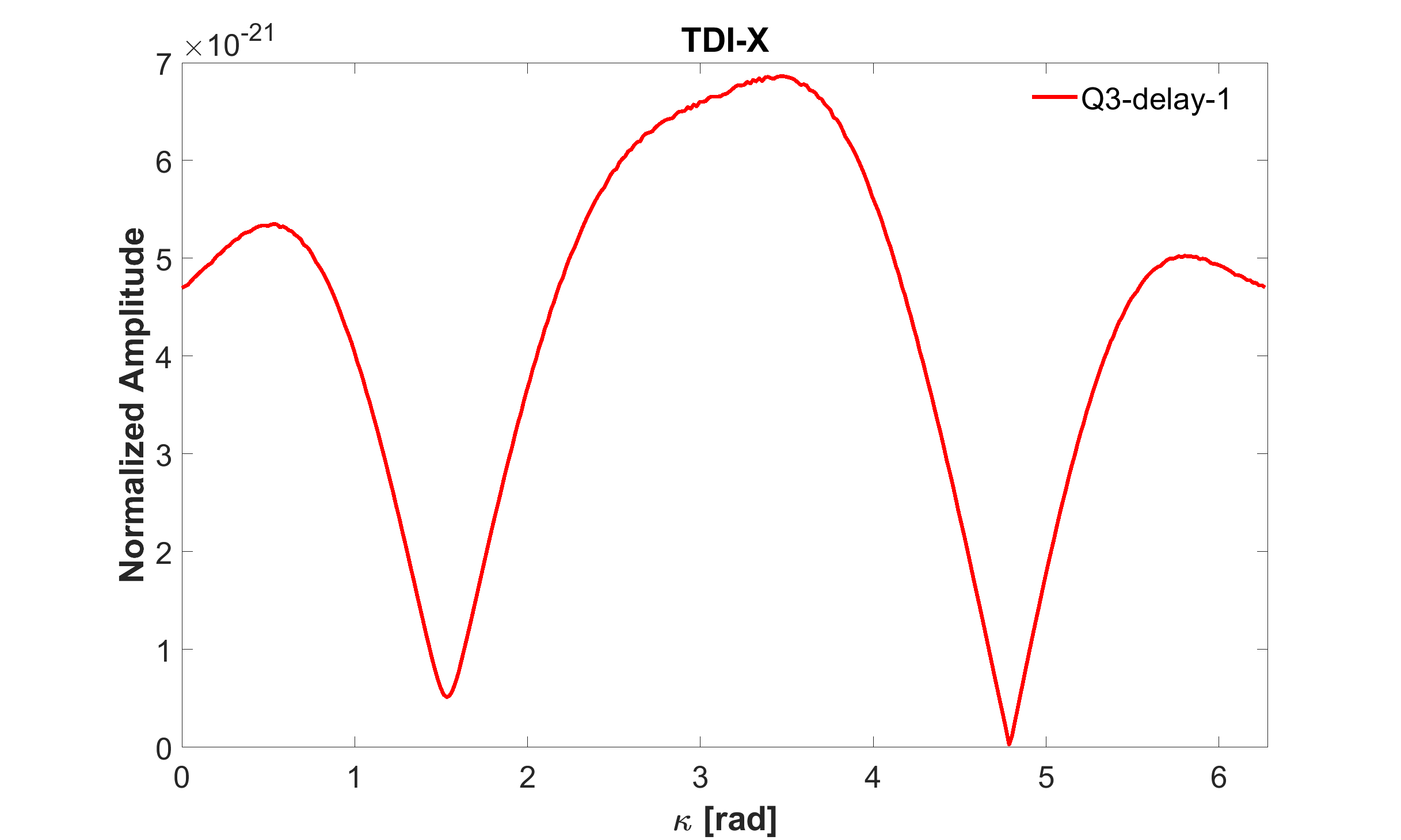}}
    \subfigure[]{\includegraphics[width=0.3\textwidth]{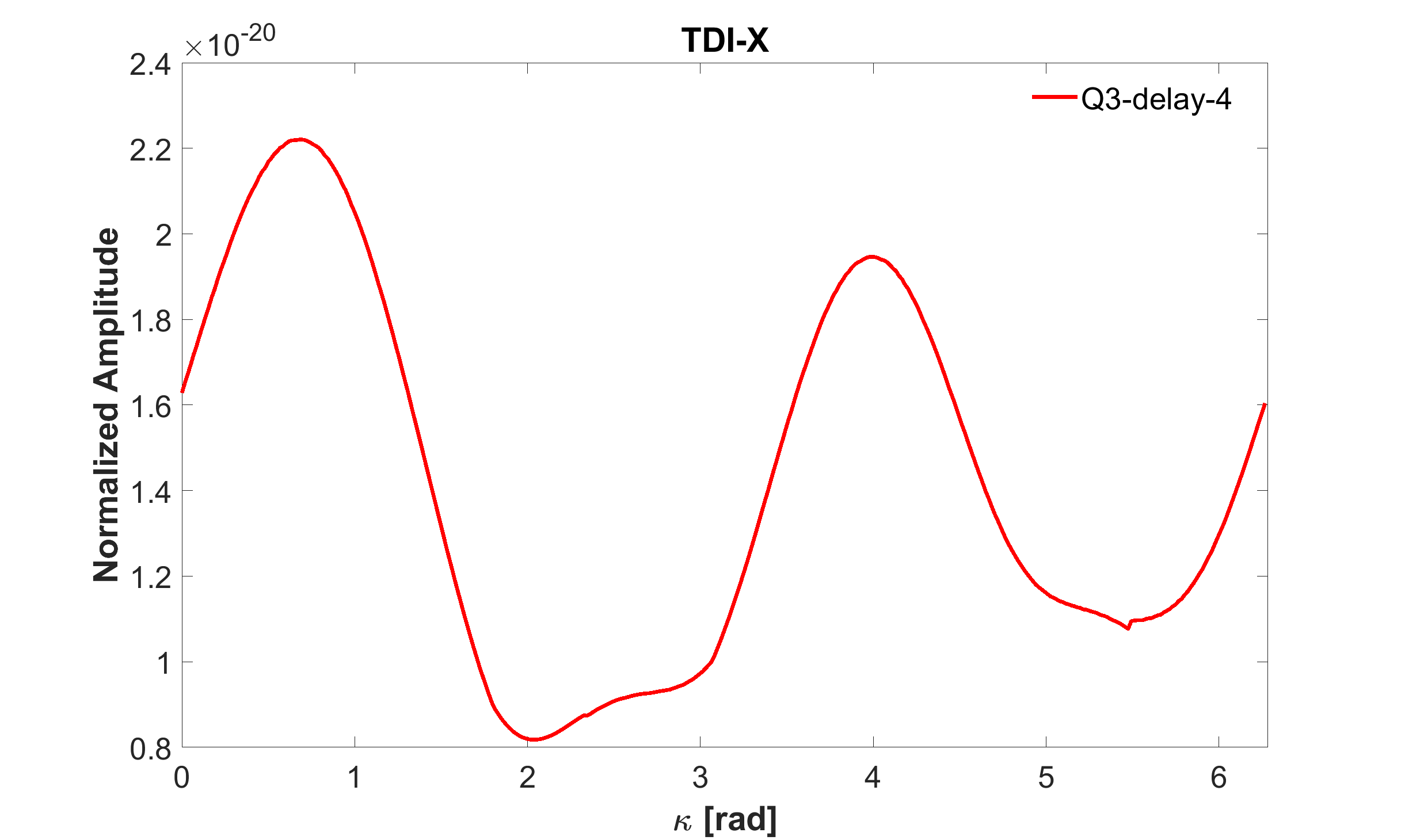}}
    \subfigure[]{\includegraphics[width=0.3\textwidth]{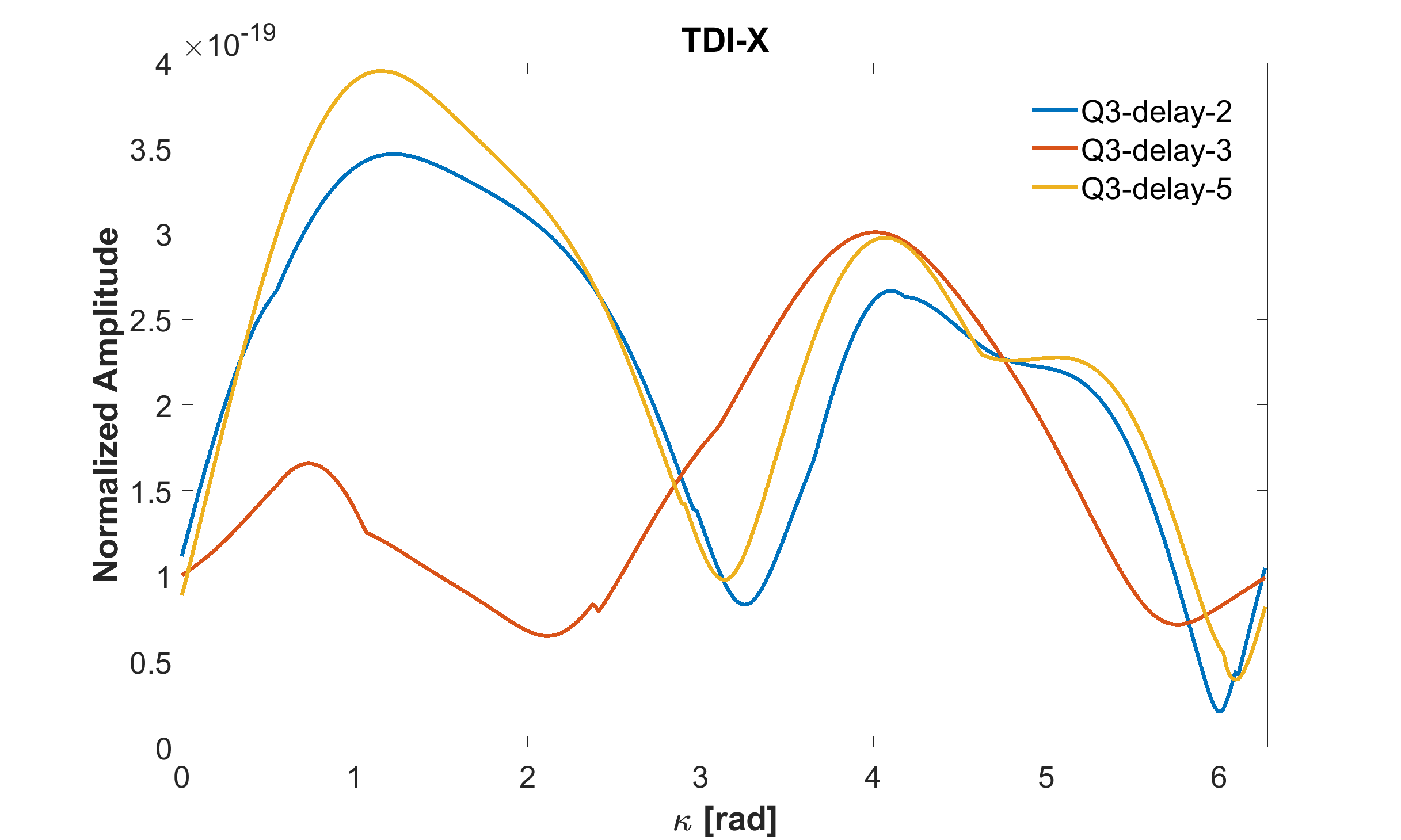}}
     \caption{The change of the normalized amplitude of the detector response TDI-X along with orbit position parameter $\kappa$ for MBHB sources: Q3-delay-1$\sim$Q3-delay-5, respectively.} 
    \label{fig:Amp_kappa_Q3-delay}
\end{figure*}

The annual modulation effect of a space-based GW detector on the source of GW is primarily a result of modulating the amplitude of the source signal. Here we select 5 MBHB sources from Q3-delay model, Q3-delay-1$\sim$Q3-delay-5 to observe the annual modulation effect of the detector on them. Here, take 365 different $\kappa$ values uniformly in the interval $\kappa \in \left[ 0,2\pi\right]$. The variation in the normalized amplitude of the detector response for the TDI-X configuration, in conjunction with the orbit position parameter $\kappa$ for each of the MBHB sources from Q3-delay-1$\sim$ Q3-delay-5 is shown in Fig.~\ref{fig:Amp_kappa_Q3-delay}.

When the normalized amplitude of the detector response for the TDI-X configuration reaches its maximum, the corresponding orbit position parameter $\kappa$ is identified as the optimal observation position. Consequently, this method allows us to choose or assume the MBHB observed at the optimal position.

Based on the optimal observation positions for MBHB sources, we calculate the TDI-X signals for these five MBHB sources. By combining the responses of the five MBHB sources from model Q3-d with the responses from 10,000 DWD confusion signals, we obtain five segments of overlapped GW signal data containing MBHB signals.

\section{Sky location of MBHB}\label{Sec:SL_MBHB}

\subsection{Bayesian inference}\label{sec:Bayes}
The Bayesian inference is based on 
calculating the  posterior probability distribution function (PDF) of  
the unknown parameter set 
$\boldsymbol{\theta}=\{\theta_{1},\dots,\theta_{m}\}$ in a given model,
which actually updates our state of belief from the prior PDF of $\boldsymbol{\theta}$ 
after taking into account the information provided by  the experimental data set $D$.
The posterior PDF is related to the prior PDF by  the Bayes's therom
\begin{align}\label{eq:Bayes}
p(\boldsymbol{\theta}|D)=\frac{\mathcal{L}(D|\boldsymbol\theta)\pi(\boldsymbol\theta)}{p(D)} ,
\end{align}
where $\mathcal{L}(D|\boldsymbol\theta)$ is the likelihood function,
and 
$\pi(\boldsymbol\theta)$ is the prior PDF which 
encompasses our state of knowledge on the values of the parameters before 
the observation of  the data.
The quantity $p(D)$ is the Bayesian evidence 
which is obtained by integrating the product  of the likelihood and the prior over
the whole volume of the parameter space
\begin{equation}
    p(D)=\int_{V} \mathcal{L}(D|\theta)\pi(\boldsymbol\theta) d\theta .
\end{equation}

The evidence is an important  quantity for Bayesian model comparison.It is straight forward to obtain the marginal PDFs of interested parameters 
$\{\theta_{1}, \dots, \theta_{n}\} (n<m) $ by integrating out other nuisance parameters $\{\theta_{n+1}, \dots, \theta_{m}\}$
\begin{equation}
    p(\theta_{1},\dots,\theta_{n})_{\text{marg}}=\int p(\boldsymbol\theta|D) \prod_{i=n+1}^{m}d\theta_{i} .
\end{equation}
The marginal PDF  is often used in visual presentation. 	
If there is no preferred value of $\theta_{i}$ in the allowed range ($\theta_{i,\text{min}}$, $\theta_{i,\text{max}}$),
the priors are taken as a flat distribution
\begin{align}\label{eq:priors}
\pi(\theta_{i}) \propto
\left\{
\begin{tabular}{ll}
1, &  \text{for } $\theta_{i,\text{min}}<\theta_{i}<\theta_{i,\text{max}}$
\\
0, & \text{otherwise}
\end{tabular}
\right. 
\end{align}
The likelihood function is often assumed to be Gaussian
\begin{align}
\mathcal{L}(D|\boldsymbol\theta)=
\prod_{i}
\frac{1}{\sqrt{2\pi \sigma_{i}^{2}}}
\exp\left[
	-\frac{(f_{\text{th},i}(\boldsymbol\theta)-f_{\text{exp},i})^{2}}{2\sigma_{i}^{2}}  
\right]  ,  
\end{align}
where $f_{\text{th},i}(\boldsymbol\theta)$ are 
the predicted $i$-th observable from the model which 
depends on the parameter set $\boldsymbol\theta$, 
and
$f_{\text{exp},i}$ are the ones measured by the experiment with uncertainty $\sigma_{i}$.

When specifying the form of the likelihood function, a posterior PDF is determined using the Markov Monte Carlo (MCMC) method based on the prior PDF and likelihood function.
\textcolor{black}{For millihertz GW simulation and Bayesian analysis, there is public code\cite{Wang:2022apn}.}
Specifically, this article uses the Metropolis-Hastings sampling method to obtain a posterior PDF.

\subsection{Wavelet decomposition and reconstruction}
\begin{figure}
	\centering 
    \includegraphics[width=0.45\textwidth]{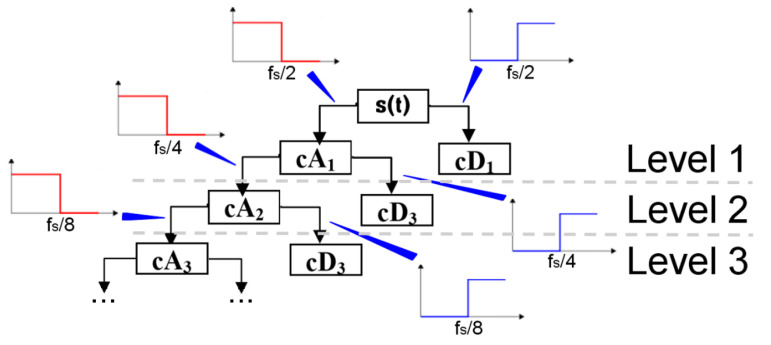}
    \caption{Wavelet decomposition process of 3 level\citep{s20247138,singh2011jpeg}. } 
    \label{fig:Wavelet_decomposition}
\end{figure}

Wavelet transform (WT) \citep{chui1992introduction, meyer1992wavelets, benedetto1993wavelets, misiti2013wavelets} is a very important and classic method in signal processing.
This paper primarily focuses on wavelet decomposition and reconstruction methods, opting for the discrete wavelet transform (DWT) for computation. Initially, the signal to be decomposed is downsampled, followed by the application of wavelet decomposition. The process involves the use of two distinct filters—namely, high-pass and low-pass filters—which are convolved with the signal to obtain the wavelet decomposition coefficients. These coefficients are categorized into two types: detail coefficients (cD), derived from high-pass filtering, and approximation coefficients (cA), derived from the low-pass filter. The specific wavelet decomposition process is illustrated in Figure~\ref{fig:Wavelet_decomposition}.
Subsequently, the detail coefficients (cD) and the approximation coefficients (cA) are selected to reconstruct the signal via wavelet reconstruction, which is considered the inverse process of wavelet decomposition.

\begin{figure*}[!htbp]
	\centering 
    \subfigure[]{\includegraphics[width=0.45\textwidth]{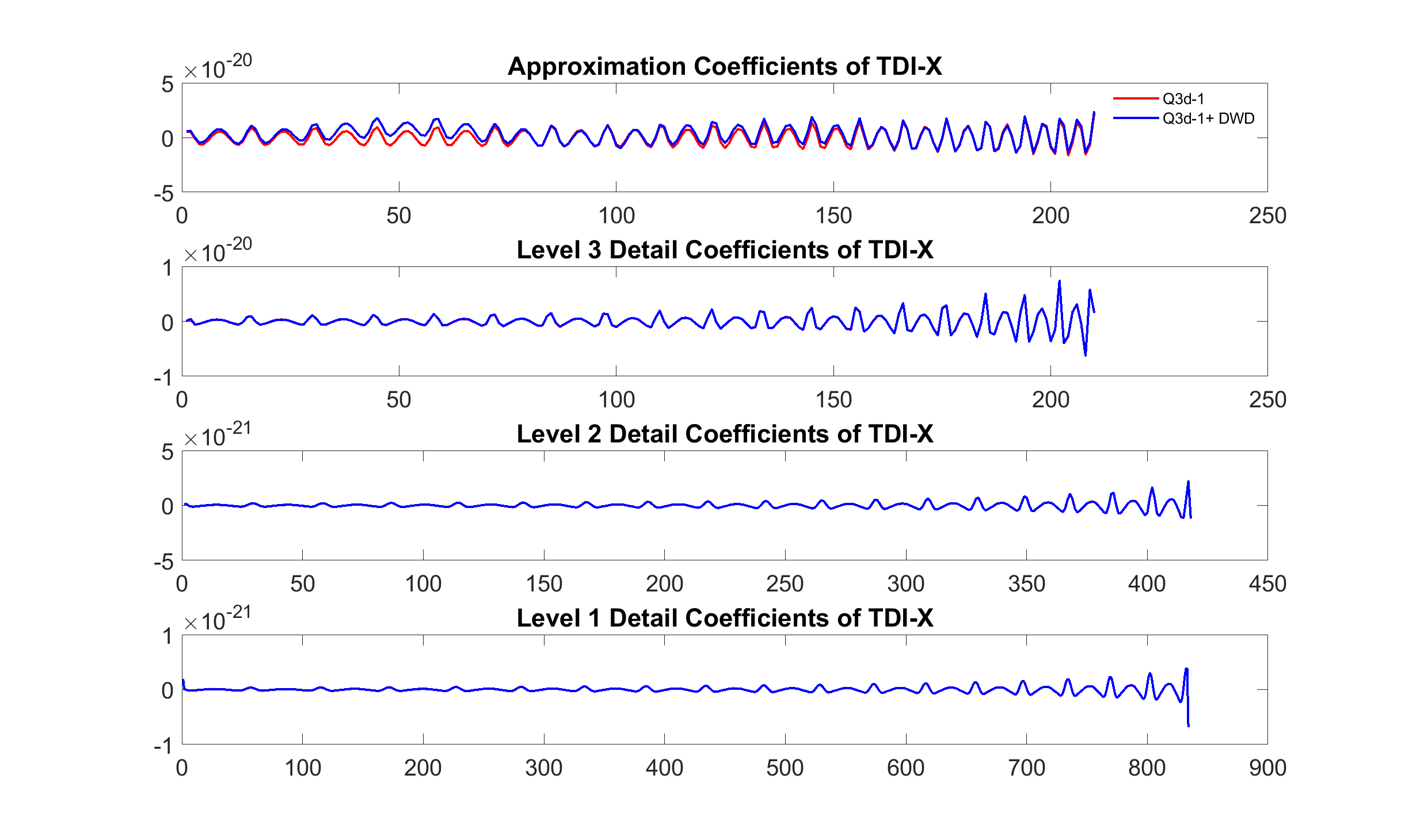}}
    \subfigure[]{\includegraphics[width=0.45\textwidth]{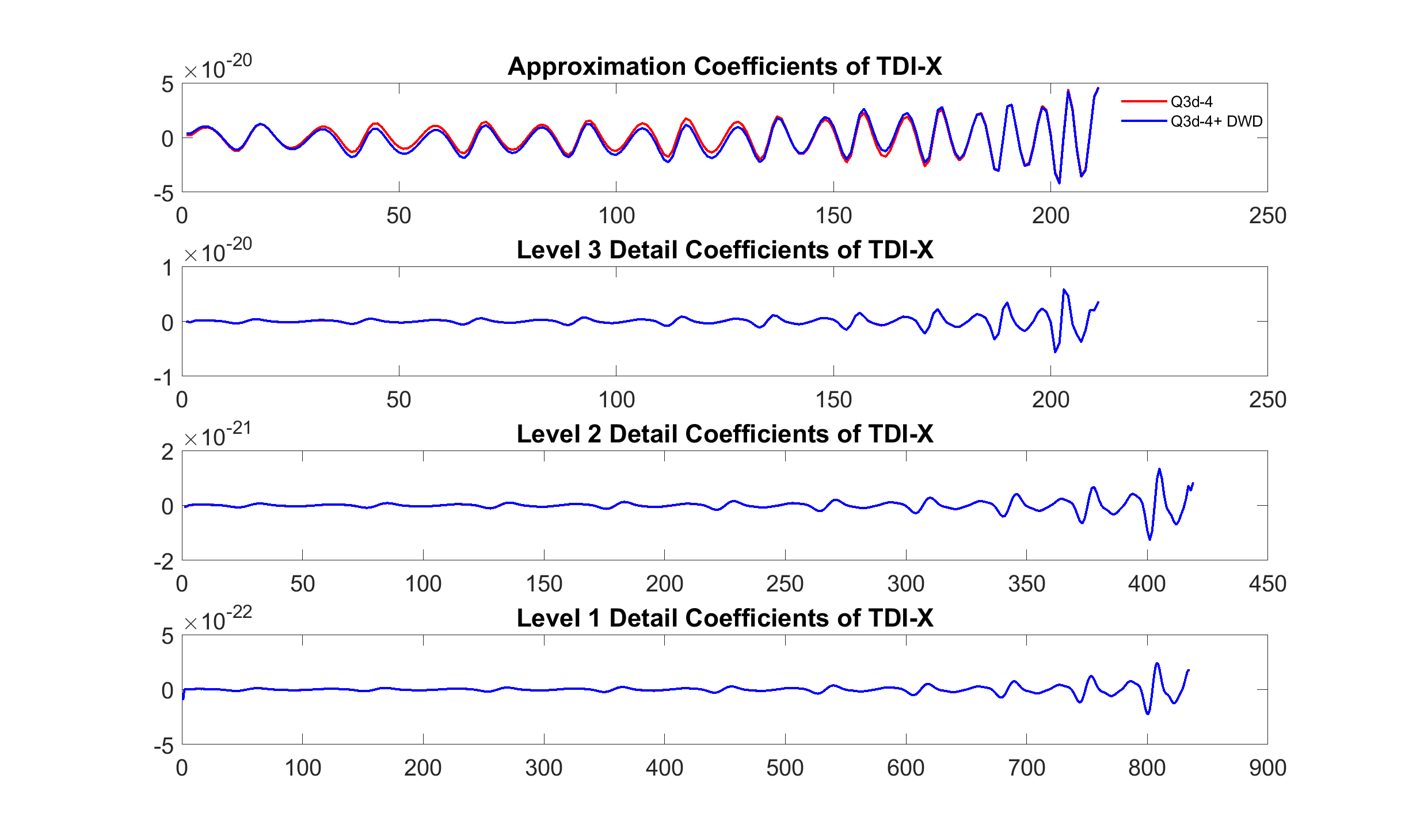}}
    \caption{Results of Q3-delay-1 and Q3-delay-4 before and after mixing with DWDs based on Wavelet decomposition at $level=3$.} 
    \label{fig:Wavelet_decomposition_Q3-delay}
\end{figure*}

Among the 5 MBHB sources identified in the Q3-delay model, we have chosen the relatively weaker sources, specifically Q3-delay-1 and Q3-delay-4, for a comparative analysis using wavelet decomposition. The decomposition is conducted at a level of 3, and the Daubechies wavelet of order 3 ($db3$) is selected for the analysis. This process yields four sets of coefficients: one set of approximate coefficients and three sets of detail coefficients. For further details, refer to Fig.~\ref{fig:Wavelet_decomposition_Q3-delay}. In the figure, the TDI-X configuration for calculating the detector response signals of each MBHB source is selected , and its wavelet decomposition coefficients are obtained.

As illustrated in Fig.~\ref{fig:Wavelet_decomposition_Q3-delay}, we observed that the primary distinction between MBHB source signal before and after the mixing with DWDs data is evident in the approximate coefficient. In contrast, the differences in the three detail coefficients are less pronounced. This phenomenon can be attributed to the nature of the coefficients: the approximate coefficient predominantly represents the lower-frequency components of the data signal, whereas the detail coefficients capture the higher-frequency components. Consequently, we proceed to filter out the approximate coefficients and reconstruct the data using only the detail coefficients, thereby obtaining new data following the wavelet transform reconstruction process.

\subsection{Posterior}

The MBHB source position parameters are set by $\boldsymbol\theta=\{ sin \beta, \lambda,  D_L\}$. The prior is chosen to be the region near the true parameter values of the five sources to be observed. Considering the program's running time, for this experiment, we selected a single Markov Chain Monte Carlo (MCMC) chain with a sample size of $136000$. The log-likelihood value is calculated for each set of samples. Subsequently, the posterior distribution of the parameters is derived using the Metropolis-Hastings sampling method.

In this paper, three sets of comparative analyses are conducted. The first set pertains to an ideal-case scenario, where it is hypothesized that a 100\% removal of the 10 kHz loud DWDs is feasible. This allows for the direct solution of the independent sources Q3-delay-1$\sim$Q3-delay-5 by applying Fast Fourier Transform (FFT) to their response data from the TDI-X configuration , which is abbreviated as IdealCase. The second set corresponds to a worst-case scenario, where no DWDs are removed, necessitating the resolution of the overlapped gravitational wave (GW) signals of Q3-delay-1$\sim$Q3-delay-5 and DWDs by directly applying FFT to their response TDI-X data, abbreviated as WorstCase. The third set also represents a worst-case scenario, which involves performing wavelet decomposition on the response data of the overlapped GW signals for Q3-delay-1 and Q3-delay-4 , along with the DWDs. Subsequently, an FFT is applied to the newly reconstructed data after filtering out the approximate coefficients, which is abbreviated as WorstCase(DWT).
Here, $level=6$ and $db6$ is selected for wavelet type of DWT.

Here, taking the $\chi^2$ logarithm of the likelihood function gives the following form:
\begin{align}\label{eq:Chisquare}
\chi^2 = \frac{(h_{\text{th},i}(\boldsymbol\theta)-h_{\text{exp},i})^{2}}{\sigma_{i}^{2}},  
\end{align}
The projected data obtained from the GW source to be solved at the best observation position is selected as the experimental data $h_{\text{exp},i}$=$h_{MBHB,i}$ for IdealCase and $h_{\text{exp},i}$=$h_{MBHB,i}+h_{DWDs}$ for WorstCase. The uncertainty of the experimental data $\sigma_{i}$=$1/\rm{SNR_{MBHB}} \times h_{MBHB,i}$ for IdealCase and $\sigma_{i}$=$1/\rm{SNR_{MBHB}} \times h_{MBHB,i}$+ $1/\rm{SNR_{DWDs}}\times h_{DWDs,i}$ for WorstCase, where $\rm SNR_{MBHB}$ and $\rm SNR_{DWDs}$ are calculated by the Eq.~(\ref{eq:SNR}) through the projected data obtained $h_{MBHB,i}$ and $h_{DWDs,i}$   
and the analytic model for the TDI-X sensitivity curve $S_n$, specifically formula (3)\citep{Wang:2020vkg} for IdealCase, adding formula (14)\citep{robson2019construction} to formula (3)\citep{Wang:2020vkg} for WorstCase.
The observation time $T_{OB}=5\times10^5 s$. See details in the  Table~\ref{tab:SNR_MCMC}.
\begin{align}\label{eq:SNR}
\rm SNR_{MBHB,DWDs} = 2\sqrt{\int \frac{|h_{MBHB,DWDs}^{2}(f)|}{S_{n}(f)}df},  
\end{align}
\begin{table*}[!htbp]
    \centering
    \footnotesize
    \setlength{\tabcolsep}{10pt}
    \renewcommand{\arraystretch}{1.5}
    \begin{tabular}{lccccc}
        \hline
        SNR & Q3-delay-1 & Q3-delay-2 & Q3-delay-3 & Q3-delay-4 & Q3-delay-5 \\
        \hline
        SNR$_{MBHB}$(IdealCase) & 15.9 & 258 & 265 & 49 & 425\\
        SNR$_{MBHB}$(WorstCase) & 15.8 & 122 & 147 & 48 & 319 \\
        SNR$_{DWDs}$(WorstCase) & 1.76 & 2.1 & 1.6 & 3.2 & 2.2 \\
        \hline            
\end{tabular}
    \caption{SNR$_{MBHB}$ and SNR$_{DWDs}$ in the ideal case and worst case for MCMC of Q3-d sources.}
    \label{tab:SNR_MCMC}
\end{table*} 

\begin{figure*}[!htbp]
	\centering 
    \subfigure[IdealCase:Q3-delay-2]{\includegraphics[width=0.3\textwidth]{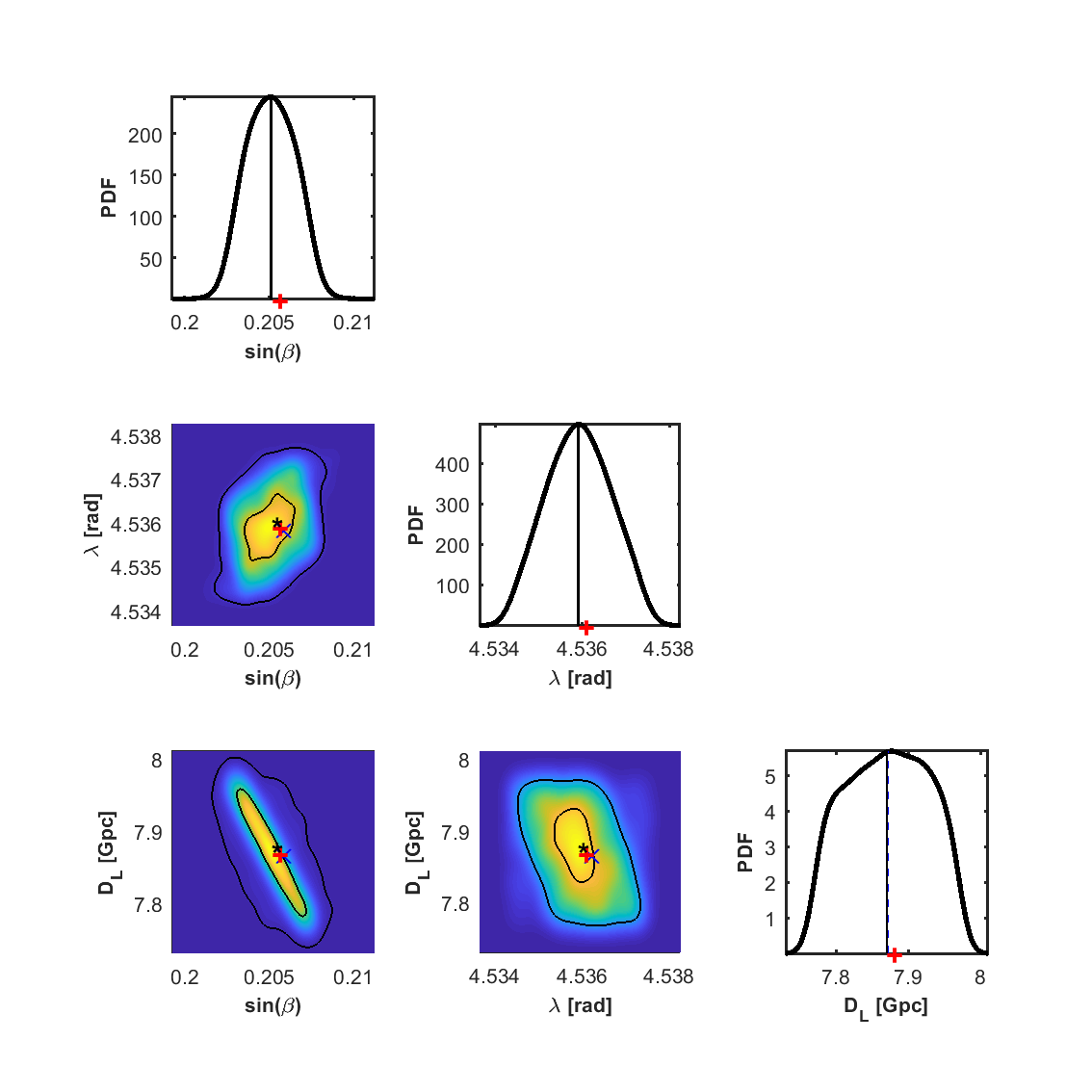}}
    \subfigure[IdealCase:Q3-delay-3]{\includegraphics[width=0.3\textwidth]{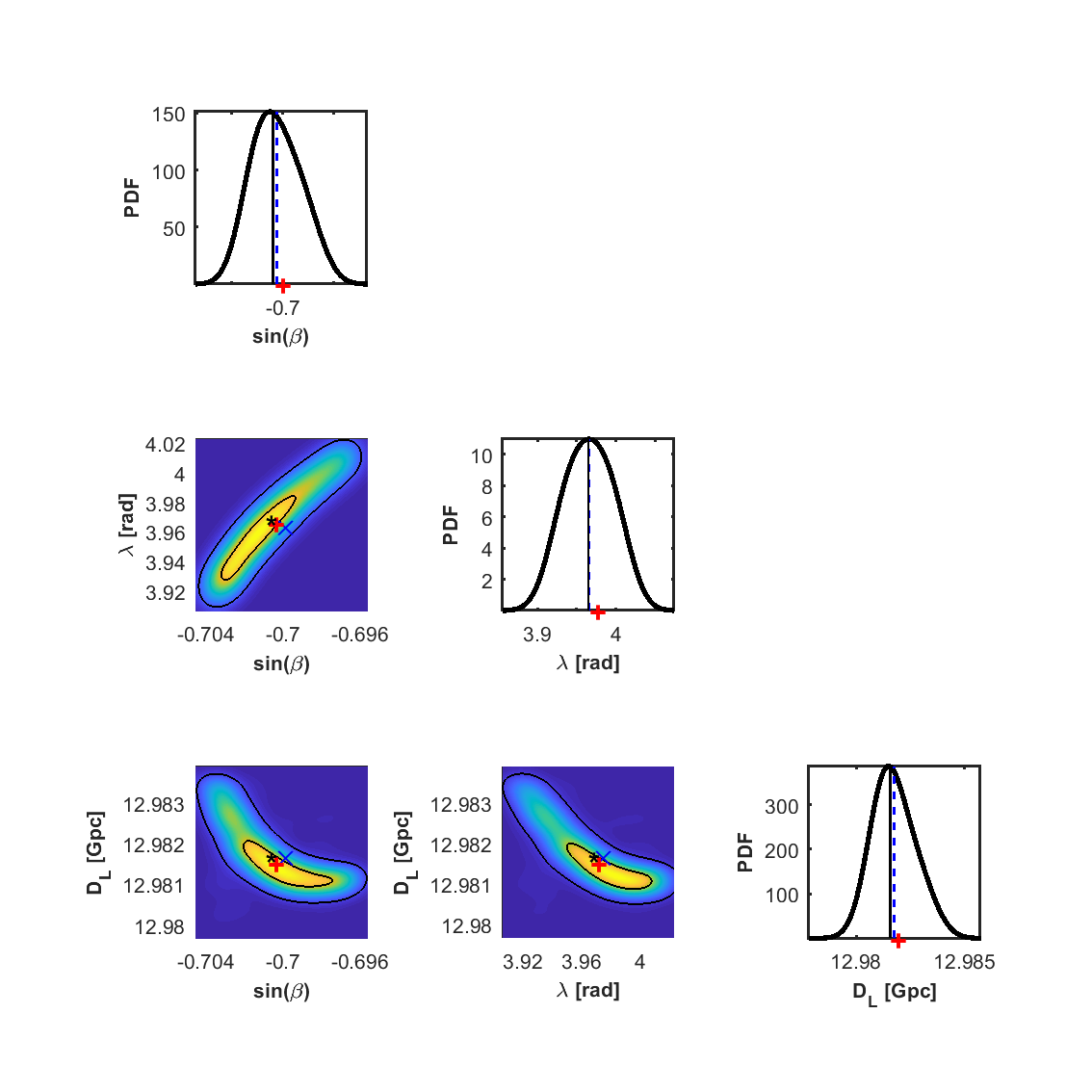}}
    \subfigure[IdealCase:Q3-delay-5]{\includegraphics[width=0.3\textwidth]{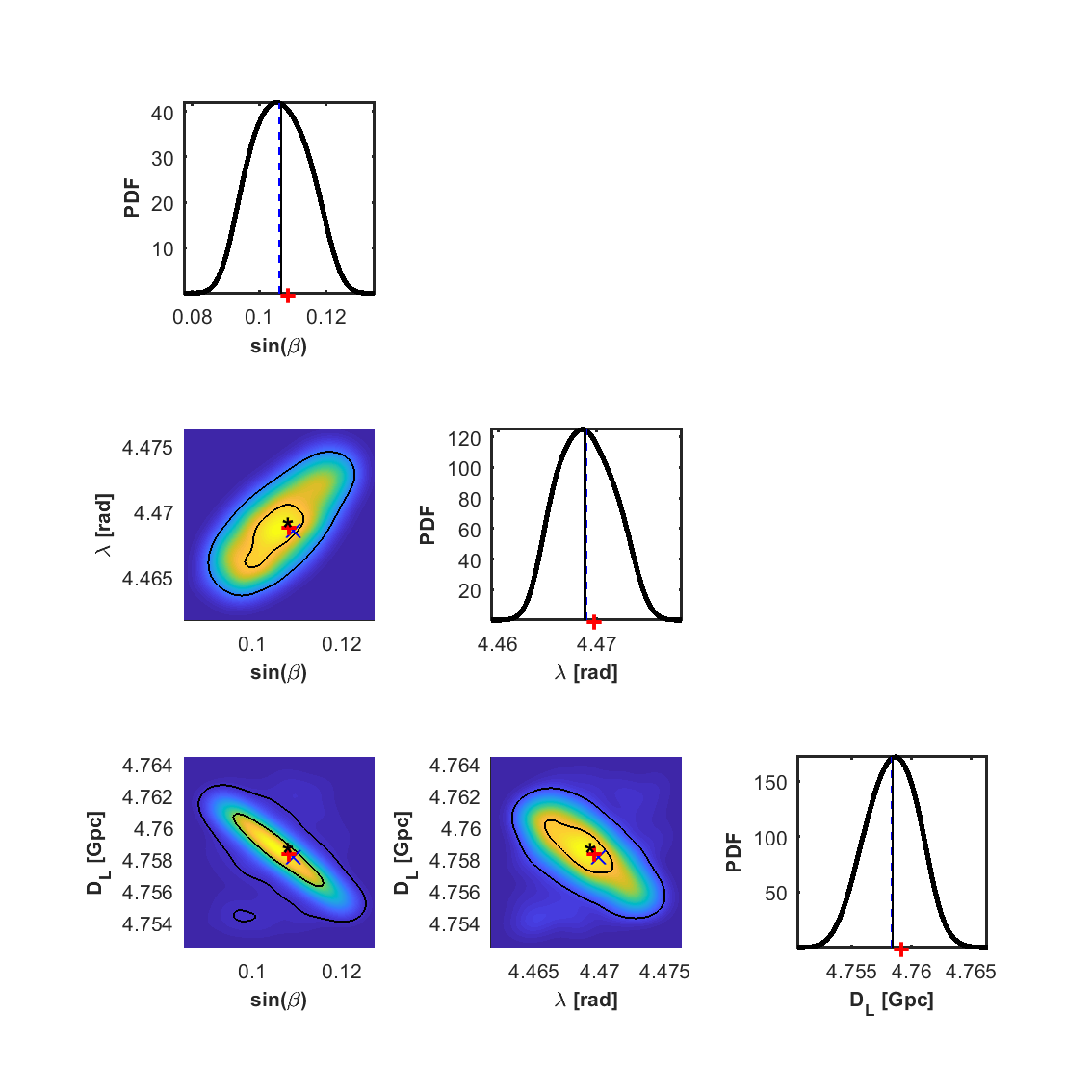}}
    \\
    \subfigure[WorstCase:Q3-delay-2]{\includegraphics[width=0.3\textwidth]{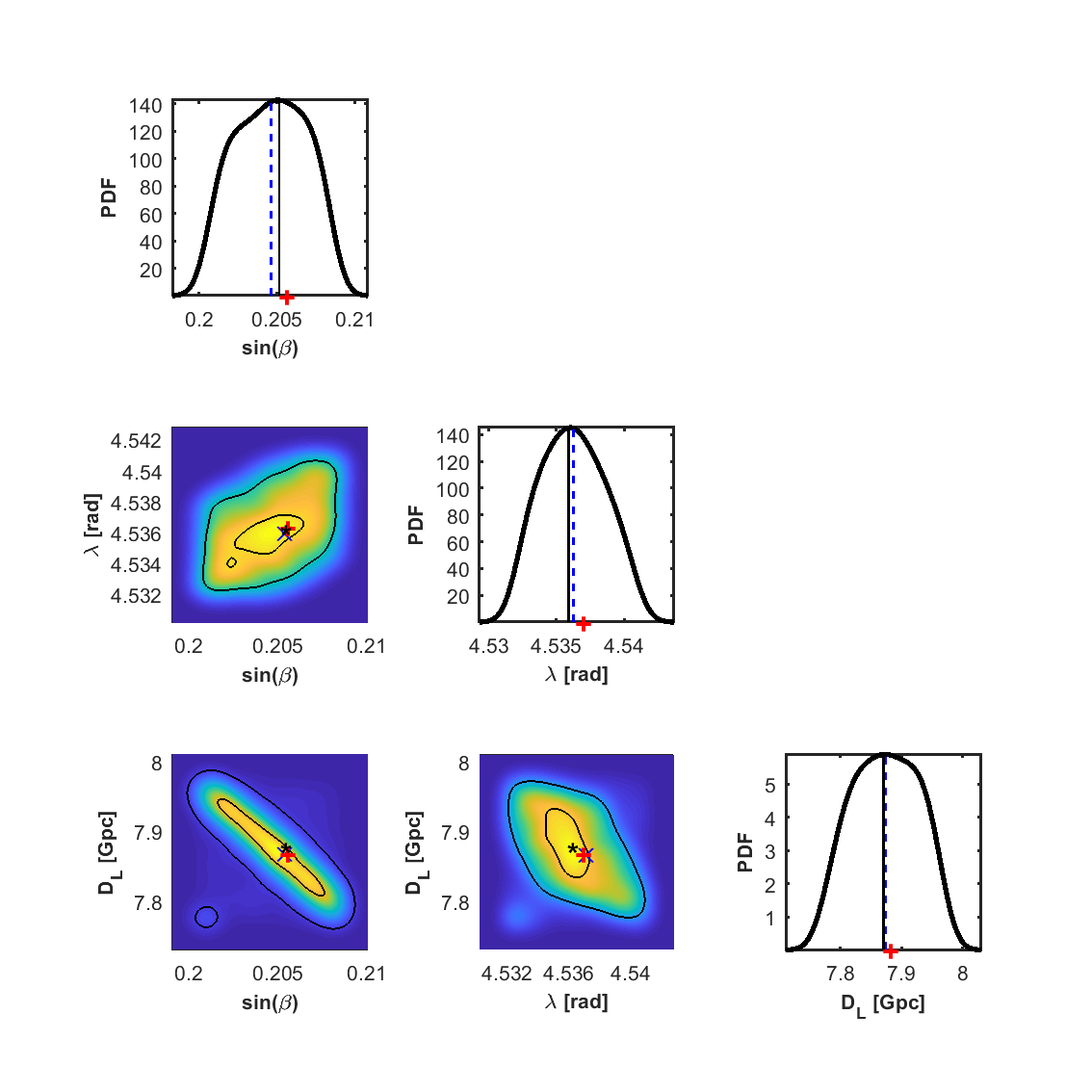}}
    \subfigure[WorstCase:Q3-delay-3]{\includegraphics[width=0.3\textwidth]{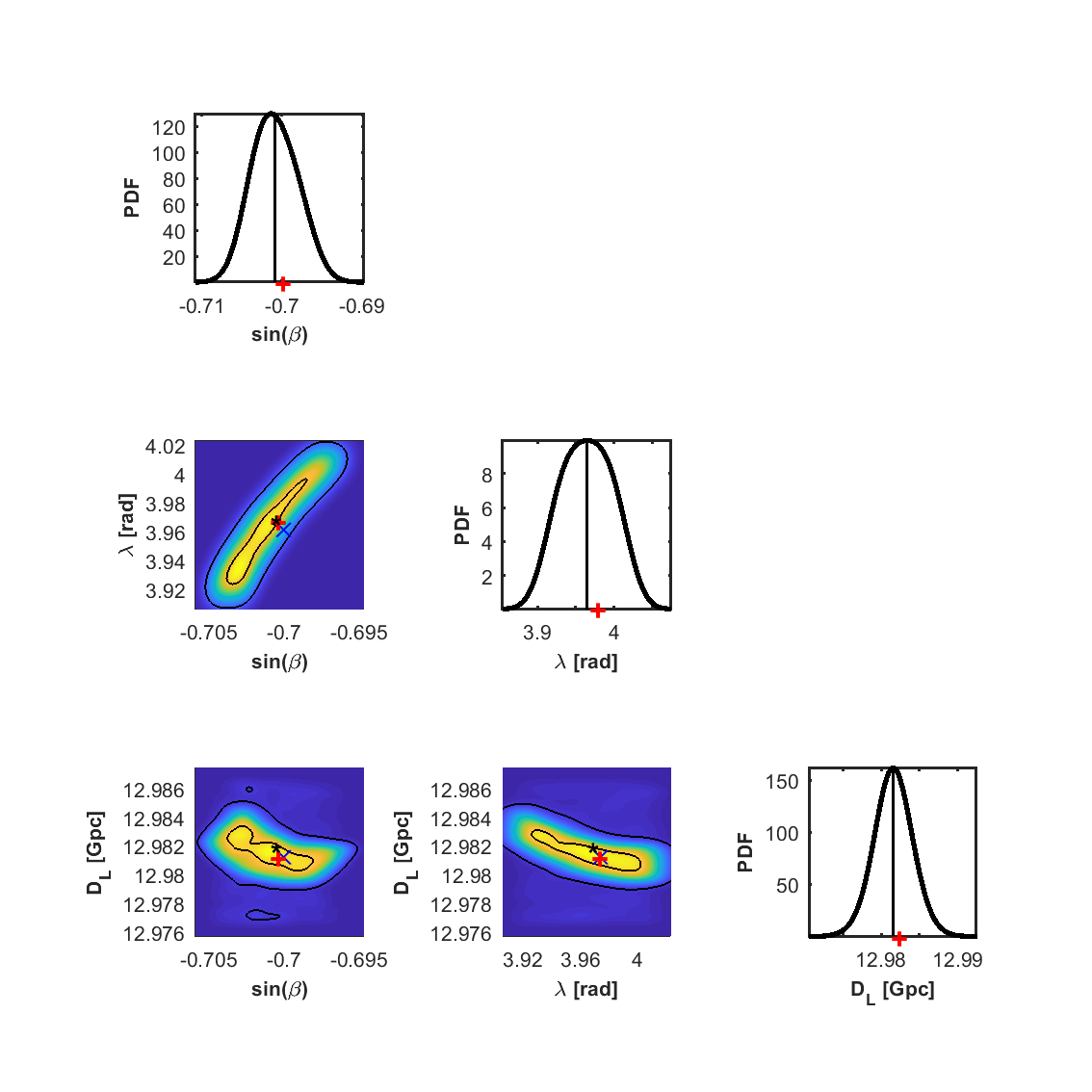}}
    \subfigure[WorstCase:Q3-delay-5]{\includegraphics[width=0.3\textwidth]{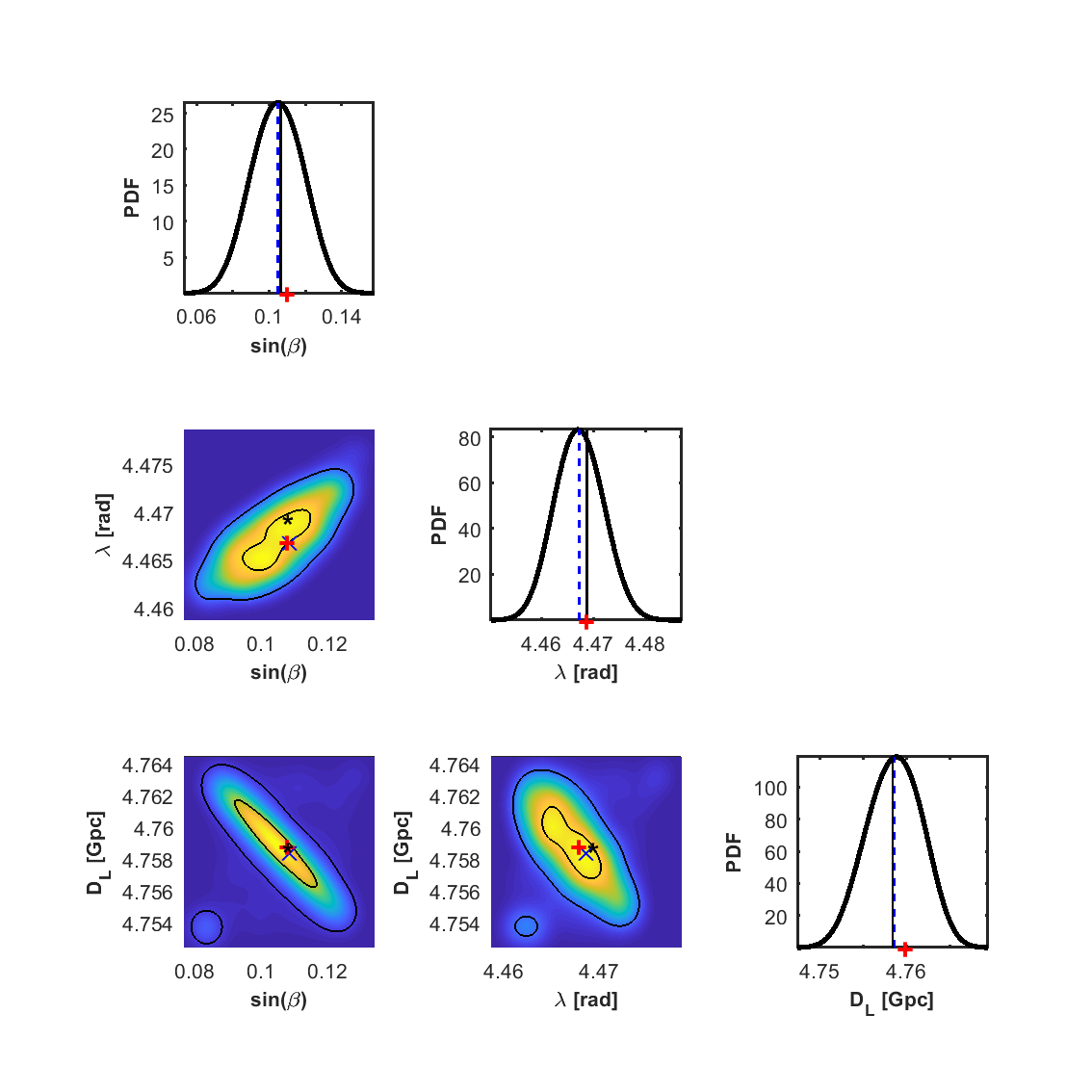}}
    \caption{The contrast of posterior distribution of the sky position parameters ($sin\beta, \lambda,  D_L$) of IdealCase and WorstCase for Q3-delay-2, Q3-delay-3 and Q3-delay-5. The black (blue) vertical line in the post-pdf of each one-dimensional marginal posterior PDF represents the true value (statistical mean). The black star (blue cross) in the pdf after each one-dimensional marginal posterior PDF represents the true value (statistical mean). The red plus sign in each one-dimensional and one-dimensional marginal posterior PDF indicates the best fit value. The contours contain the $68\%$ and $95\%$ probability regions of the parameter estimates.}
    \label{fig:Q3-d-MBHB-DWDs}%
\end{figure*}
  
\begin{figure*}[!htbp]
	\centering 
    \subfigure[IdealCase:Q3-delay-1]{\includegraphics[width=0.3\textwidth]{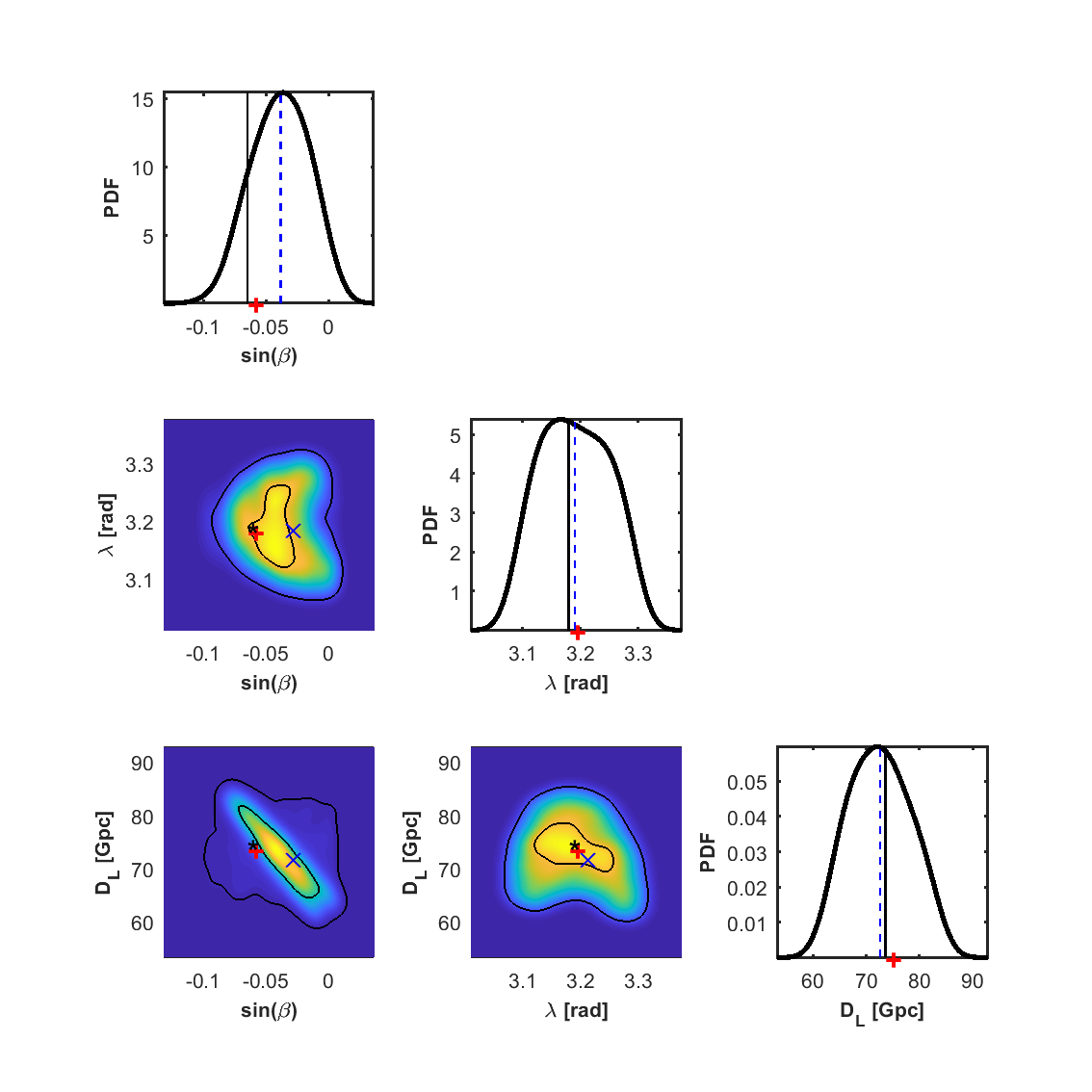}}
    \subfigure[WorstCase:Q3-delay-1]{\includegraphics[width=0.3\textwidth]{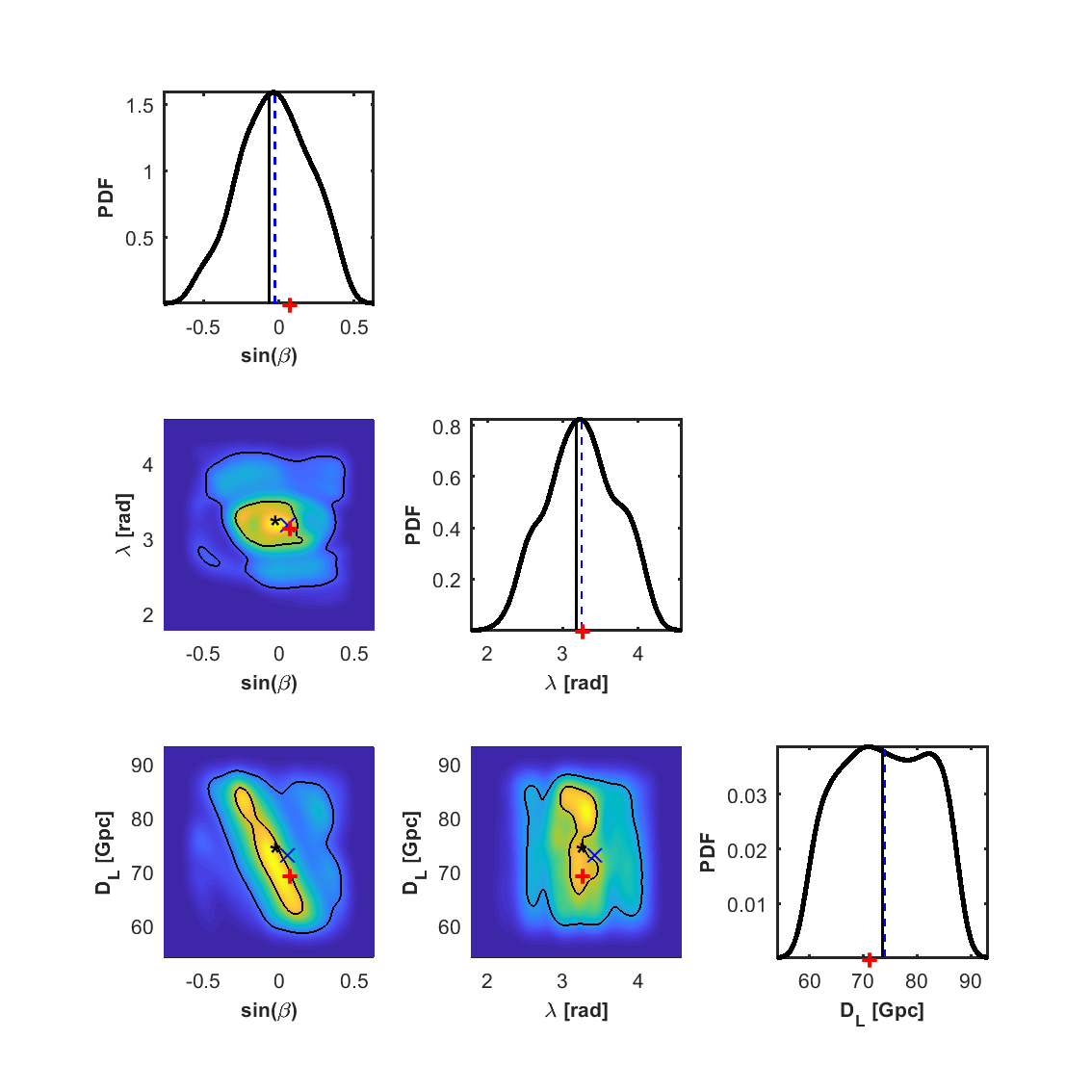}}
    \subfigure[WorstCase(DWT):Q3-delay-1]{\includegraphics[width=0.3\textwidth]{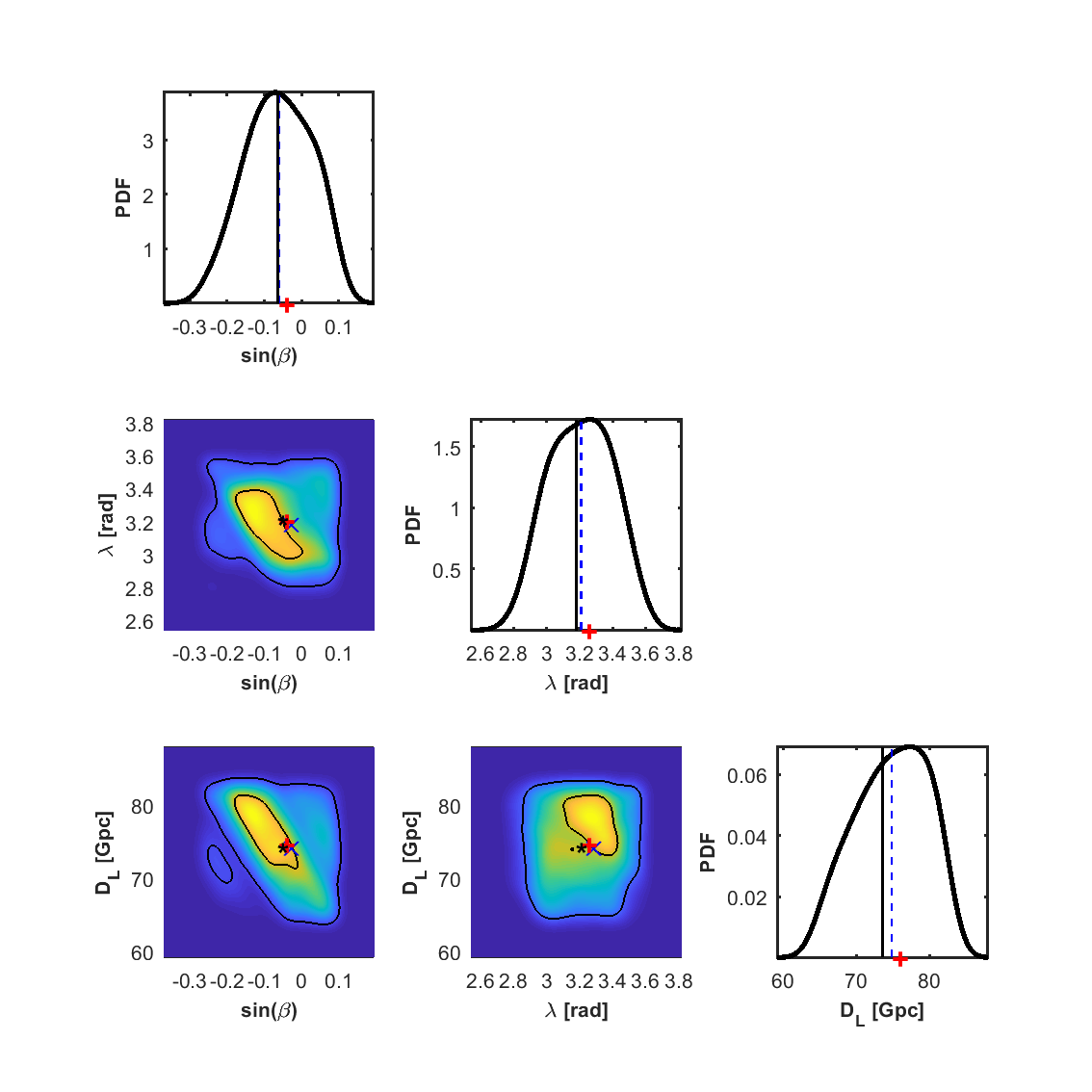}}
    \\
    \subfigure[IdealCase:Q3-delay-4]{\includegraphics[width=0.3\textwidth]{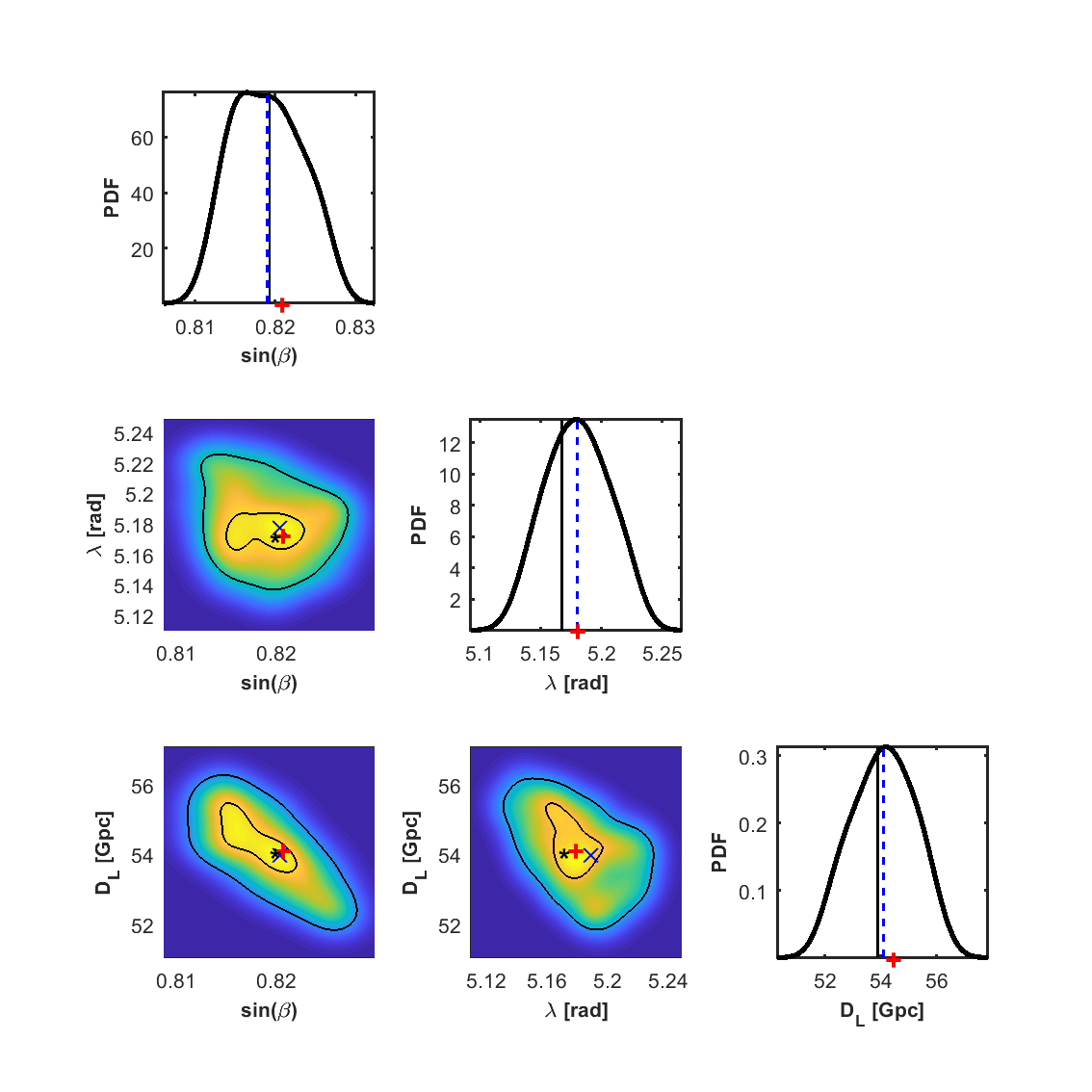}}
    \subfigure[WorstCase:Q3-delay-4]{\includegraphics[width=0.3\textwidth]{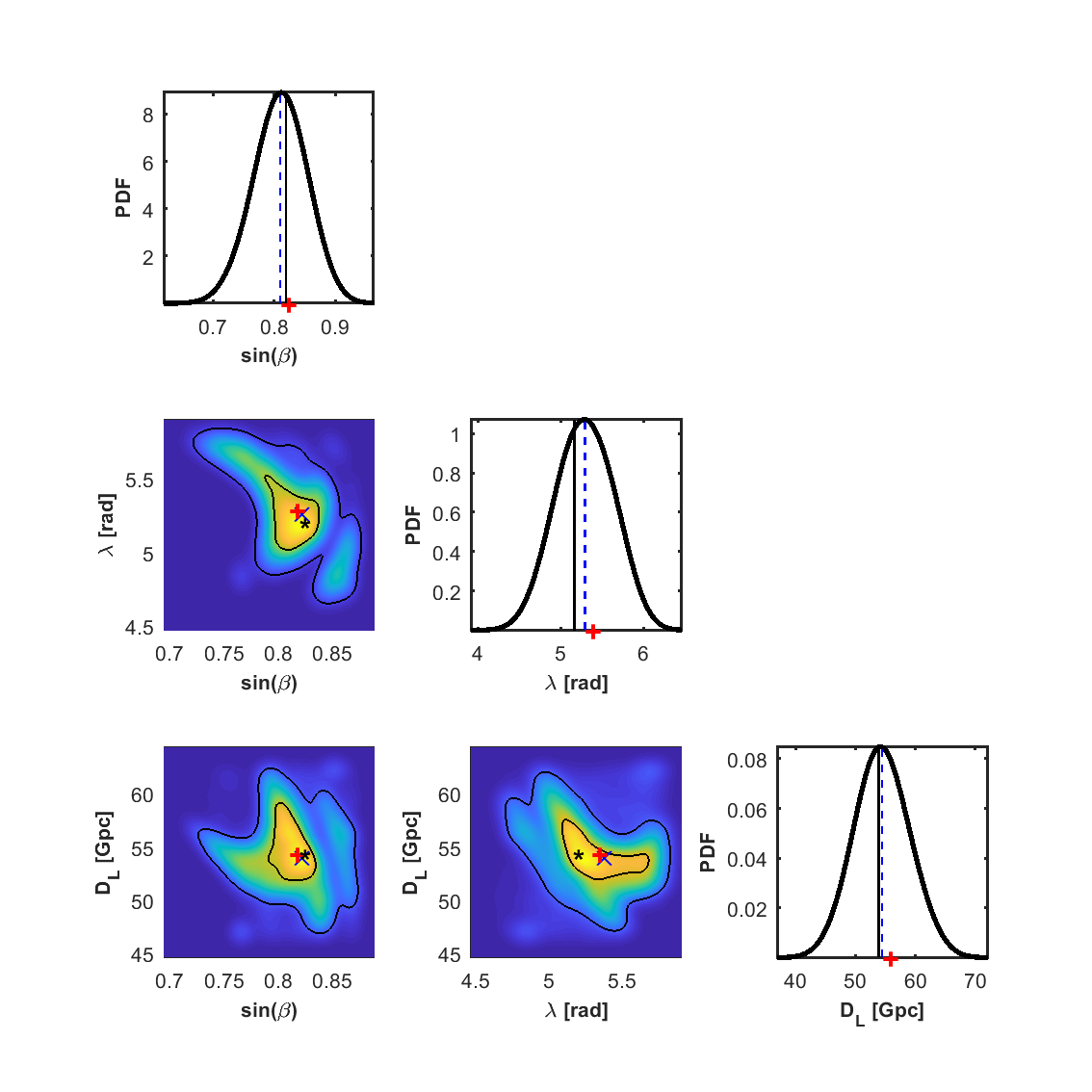}}
    \subfigure[WorstCase(DWT):Q3-delay-4]{\includegraphics[width=0.3\textwidth]{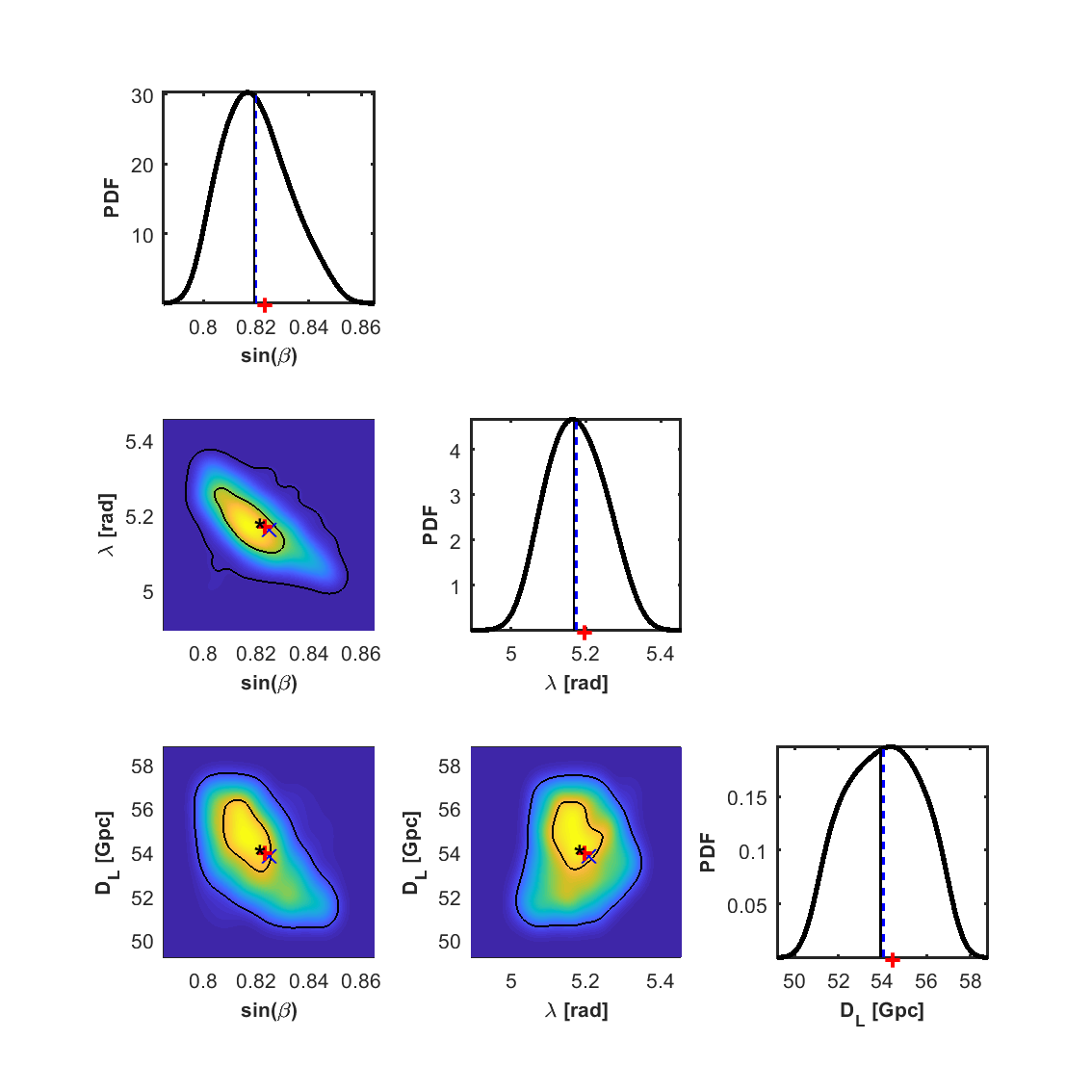}}
    \caption{The contrast of posterior distribution of the sky position parameters of IdealCase, WorstCase and WorstCase(DWT) for Q3-delay-1 and Q3-delay-4.The black (blue) vertical line in the post-pdf of each one-dimensional marginal posterior PDF represents the true value (statistical mean). The black star (blue cross) in the pdf after each one-dimensional marginal posterior PDF represents the true value (statistical mean). The red plus sign in each one-dimensional and one-dimensional marginal posterior PDF indicates the best fit value. The contours contain the $68\%$ and $95\%$ probability regions of the parameter estimates.}
    \label{fig:Q3-d-MBHB-DWDs-DWT}%
\end{figure*}

The specific posterior distributions for the first and second sets of Q3-delay-2, Q3-delay-3, and Q3-delay-5 are shown in Figure~\ref{fig:Q3-d-MBHB-DWDs}. The specific posterior distributions for the first, second, and third sets of Q3-delay-1 and Q3-delay-4 are depicted in Figure~\ref{fig:Q3-d-MBHB-DWDs-DWT}.
These figures present the one-dimensional and two-dimensional marginal posterior probability density distributions of the sky position parameters $\sin\beta, \lambda,$ and $ D_L $ of the source. The best fitting values and statistical averages are derived from these distributions. The contour lines demarcate the 68\% and 95\% probability regions of the parameter estimates. In the posterior probability density function (post-PDF) plots for each one-dimensional marginal posterior, the black (blue) vertical line represents the true value (statistical mean). Additionally, the black star (blue cross) in each plot signifies the true value (statistical mean). The red plus sign in each one-dimensional marginal posterior plot indicates the best fit value. The contours encapsulate the 68\% and 95\% probability regions of the parameter estimates.

The error in the solid angle\cite{Blaut:2011zz} is given by,  
\begin{equation}
    \Delta \Omega_s = 2\pi\cos{\beta}\sqrt{C_{\beta\beta}C_{\lambda\lambda}-C_{\beta\lambda}^2}
\end{equation}
The covariance matrix C is calculated from the posterior distribution of the parameters $\beta$ and $\lambda$. 

Table~\ref{tab:Q3d_ParametersEstimation} shows the results of parameter estimation of Q3-delay-1$\sim$Q3-delay-5. The sky location parameters and corresponding true values, the confidence interval at 68$\%$ and 95$\%$ confidence level, luminosity distance uncertainty, $\Delta D_L$/$D_L$ based on 95$\%$ confidence level and angular resolution $\Delta \Omega_s$ are shown in this table.

\begin{table*}[!htbp]
    \centering
    \footnotesize
    \setlength{\tabcolsep}{6pt}
    \renewcommand{\arraystretch}{1.3}
    \begin{tabular}{|l|c|cc|cc|c|}
        \hline
        Q3-d & three sets of comparisons & parameter & real value &   95$\%$ CL & lg($\Delta D_L$/$D_L$) & lg($\Delta \Omega_s [deg^2]$) \\
        \hline
        \multirow{9}*{Q3-delay-1} & \multirow{3}*{IdealCase}  & D$_L$ [Gpc] & {73.58} & [61.42,  83.76] & \multirow{3}*{-0.5175} & \multirow{3}*{1.398} \\
                & \multirow{3}*{}  &  sin$\beta$ & -0.06466 & [-0.08449,  0.007505]] & \multirow{3}*{} & \multirow{3}*{} \\
                & \multirow{3}*{}  &  $\lambda$[deg]& 182.2 & [176.4,  189.2]  & \multirow{3}*{} & \multirow{3}*{} \\  
        \cline{2-7}
           & \multirow{3}*{WorstCase}  & D$_L$ [Gpc] & {73.58} &  [60.04,  87.9]  & \multirow{3}*{-0.4219} & \multirow{3}*{3.330} \\
                & \multirow{3}*{}  &  sin$\beta$ & -0.06466 & [-0.4802,  0.4238] & \multirow{3}*{} & \multirow{3}*{} \\
                & \multirow{3}*{}  &  $\lambda$[deg]& 182.2 & [136.1,  236.4] & \multirow{3}*{} & \multirow{3}*{} \\  
        \cline{2-7}
            & \multirow{3}*{WorstCase(DWT)}  & D$_L$ [Gpc] & {73.58} &  [84.56, 62.18]  & \multirow{3}*{-0.5169} & \multirow{3}*{2.494} \\
                & \multirow{3}*{}  &  sin$\beta$ & -0.06466 &  [-0.2414,  0.1186]  & \multirow{3}*{} & \multirow{3}*{} \\
                & \multirow{3}*{}  &  $\lambda$[deg]& 182.2 &  [162.6,  205]   & \multirow{3}*{} & \multirow{3}*{} \\  
         \hline
         \multirow{6}*{Q3-delay-2} & \multirow{3}*{IdealCase} & D$_L$ [Gpc] & {7.870} & [7.772,  7.97] & \multirow{3}*{-1.599} & \multirow{3}*{-1.734} \\
                & \multirow{3}*{}  &  sin$\beta$ & 0.2051 & [0.2024,  0.2078] & \multirow{3}*{} & \multirow{3}*{} \\
                & \multirow{3}*{}  &  $\lambda$[deg]& 259.9 & [259.8,  260] & \multirow{3}*{} & \multirow{3}*{} \\  
        \cline{2-7}
           & \multirow{3}*{WorstCase}  & D$_L$ [Gpc] & {7.870} & [7.773,  7.974]  & \multirow{3}*{-1.591} & \multirow{3}*{-1.033} \\
                & \multirow{3}*{}  &  sin$\beta$ & 0.2051 & [0.2004,  0.2088] & \multirow{3}*{} & \multirow{3}*{} \\
                & \multirow{3}*{}  &  $\lambda$[deg]& 259.9 & [259.6,  260.2]  & \multirow{3}*{} & \multirow{3}*{} \\  
         \hline
         \multirow{6}*{Q3-delay-3} & \multirow{3}*{IdealCase} & D$_L$ [Gpc] & {12.98} & [12.98,  12.98] & \multirow{3}*{-3.49} & \multirow{3}*{-0.9463} \\         
                & \multirow{3}*{}  &  sin$\beta$ & -0.7009 & [-0.7054,  -0.6958] & \multirow{3}*{} & \multirow{3}*{} \\
                & \multirow{3}*{}  &  $\lambda$[deg]& 227.2 & [223.6,  230.8] & \multirow{3}*{} & \multirow{3}*{} \\  
        \cline{2-7}
           & \multirow{3}*{WorstCase}  & D$_L$ [Gpc] & {12.98} & [12.98,  12.99] &  \multirow{3}*{-3.105} & \multirow{3}*{-0.3645} \\
           & \multirow{3}*{}  &  sin$\beta$ & -0.7009 & [-0.7065,  -0.6953] & \multirow{3}*{} & \multirow{3}*{} \\
                & \multirow{3}*{}  &  $\lambda$[deg]& 227.2 & [223.4,  230.9]  & \multirow{3}*{} & \multirow{3}*{} \\  
         \hline
         \multirow{9}*{Q3-delay-4} & \multirow{3}*{IdealCase} & D$_L$ [Gpc] & {53.90} & [51.9,  56.32] & \multirow{3}*{-1.086} &  \multirow{3}*{0.05814} \\
                & \multirow{3}*{}  &  sin$\beta$ & 0.8193 & [0.809,  0.829] & \multirow{3}*{} & \multirow{3}*{} \\
                & \multirow{3}*{}  &  $\lambda$[deg]& 296.0 & [293.4,  300.3]  & \multirow{3}*{} & \multirow{3}*{} \\  
        \cline{2-7}
           & \multirow{3}*{WorstCase}  & D$_L$ [Gpc] & {53.90} & [45.24,  63.6] & \multirow{3}*{-0.4677} & \multirow{3}*{1.890} \\
                & \multirow{3}*{}  &  sin$\beta$ & 0.8193 & [0.7198,  0.8998] & \multirow{3}*{} & \multirow{3}*{} \\
                & \multirow{3}*{}  &  $\lambda$[deg]& 296.0 & [263.6,  342.7] & \multirow{3}*{} & \multirow{3}*{} \\  
        \cline{2-7}
            & \multirow{3}*{WorstCase(DWT)}  & D$_L$ [Gpc] & {53.90} & [50.95,  57.16] & \multirow{3}*{-0.9387} & \multirow{3}*{0.8350} \\
                & \multirow{3}*{}  &  sin$\beta$ & 0.8193 & [0.795,  0.845] & \multirow{3}*{} & \multirow{3}*{} \\
                & \multirow{3}*{}  &  $\lambda$[deg] & 296.0 & [287.9,  305.1] & \multirow{3}*{} & \multirow{3}*{} \\  
                \hline
         \multirow{6}*{Q3-delay-5} & \multirow{3}*{IdealCase} & D$_L$ [Gpc] & {4.758} & [4.754,  4.763] & \multirow{3}*{-2.753} &  \multirow{3}*{-0.5602} \\
                & \multirow{3}*{}  &  sin$\beta$ & 0.1065 & [0.09038,  0.1218] & \multirow{3}*{} & \multirow{3}*{} \\
                & \multirow{3}*{}  &  $\lambda$[deg]& 256 & [255.7,  256.4] & \multirow{3}*{} & \multirow{3}*{} \\  
        \cline{2-7}
           & \multirow{3}*{WorstCase}  & D$_L$ [Gpc] & {4.758} & [4.752,  4.765] & \multirow{3}*{-2.584} & \multirow{3}*{-0.3234} \\
                & \multirow{3}*{}  &  sin$\beta$ &  0.1065 & [0.07702,  0.1326] & \multirow{3}*{} & \multirow{3}*{} \\
                & \multirow{3}*{}  &  $\lambda$[deg]& 256 & [255.4,  256.5] & \multirow{3}*{} & \multirow{3}*{} \\  
        \hline
    \end{tabular}
    \caption{The results of three sets of comparisons of sky location parameter estimation of Q3-delay-1$\sim$Q3-delay-5. Three sets of comparisons are given in column 2. The sky location parameters and their true values are given in column 3. The confidence interval and luminosity distance uncertainty, $\Delta D_L$/$D_L$ based on 95$\%$ confidence level are given in column 4. The angular resolution $\Delta \Omega_s$ is shown in column 5.}
    \label{tab:Q3d_ParametersEstimation}
\end{table*}
As seen in Figure~\ref{fig:Q3-d-MBHB-DWDs}, Figure~\ref{fig:Q3-d-MBHB-DWDs-DWT}, and Table~\ref{tab:Q3d_ParametersEstimation}, for relatively stronger MBHB sources such as Q3-delay-2, Q3-delay-3, and Q3-delay-5, the precision of luminosity distance, represented by the ratio ($\frac{\Delta D_L}{D_L}$) at the $95\%$ confidence level, derived from the second set WorstCase are comparable to those obtained from the first set IdealCase.
However, for MBHB sources of lower intensity, such as Q3-delay-1 and Q3-delay-4, we find that the precision of luminosity distance, represented by the ratio ($\frac{\Delta D_L}{D_L}$) at the $95\%$ confidence level, and the angular resolutions, denoted by ($\Delta \Omega_s$), derived from the second set WorstCase are much worse than those from the first set IdealCase.
For MBHB sources of lower intensity, the precision of luminosity distance, represented by the ratio ($\frac{\Delta D_L}{D_L}$) at the $95\%$ confidence level, is enhanced when utilizing wavelet decomposition and reconstruction methods for the analysis of overlapping gravitational wave (GW) signals from MBHB and DWD sources, with improvements by factors of two and three, compared to those from the second set WorstCase, corresponding to Q3-delay-1 and Q3-delay-4, respectively. The angular resolution $ \Delta \Omega_s $ derived from the third set WorstCase(DWT) is enhanced by $\sim$ 20 and $\sim$ 40 times when employing wavelet decomposition and reconstruction methods, compared to those from the second set WorstCase, corresponding to Q3-delay-1 and Q3-delay-4, respectively. The wavelet decomposition and reconstruction method can significantly mitigate the impact of overlap with DWDs for MBHB sources of lower intensity.

\section{Conclusions}\label{Sec:conclusion}
The GW signals from MBHB merging linger for much longer in the detector sensitive band, which overlap with the continuous and nearly monochromatic GW signals of DWDs sources. The data analysis of overlapped GW signals becomes already a general issue to be solved. In this paper, we focus on sky location of MBHBs from overlapped GW signals with DWDs sources. 
In fact, the amplitude of GW signals is modulated by the periodic motion of GW detectors on the solar orbit. For the first time, the effects of the periodic orbit position parameter of space-based laser interferometer detectors on sky location of MBHBs are considered. 
In this paper, the TDI-X configuration from the first generation TDI combinations is chosen to calculate the detector responses.
At the optimal observation orbital position, the detector captures the merger signal from the MBHB wave source. Concurrently, the detector response signal intensity to the wave source attains its maximum.

The model Q3-d for the population of MBHBs is selected. At the optimal orbit position, observations of Q3-d sources are conducted. Location parameter estimation for the five MBHB sources is performed using the Metropolis-Hastings Markov Chain Monte Carlo (MCMC) method. In this paper, three sets of comparisons are carried out:
\begin{itemize}
\item The first set pertains to an ideal-case scenario, where it is hypothesized that a 100\% removal of the 10 kHz loud DWDs is feasible. This allows for the direct solution of the independent sources Q3-delay-1$\sim$Q3-delay-5 by applying Fast Fourier Transform (FFT) to their response data from the TDI-X configuration , which is abbreviated as IdealCase. 
\item The second set corresponds to a worst-case scenario, where no DWDs are removed, necessitating the resolution of the overlapped gravitational wave (GW) signals of Q3-delay-1$\sim$Q3-delay-5 and DWDs by directly applying FFT to their response TDI-X data, abbreviated as WorstCase.
\item The third set also represents a worst-case scenario, which involves performing wavelet decomposition on the response data of the overlapped GW signals for Q3-delay-1 and Q3-delay-4 , along with the DWDs. Subsequently, an FFT is applied to the newly reconstructed data after filtering out the approximate coefficients, which is abbreviated as WorstCase(DWT).
\end{itemize}
Here, a level of 6 and a Daubechies wavelet with 6 coefficients (db6) are selected for the discrete wavelet transform (DWT).

For the relatively stronger MBHB sources, such as Q3-delay-2, Q3-delay-3, and Q3-delay-5, when comparing IdealCase with the second set WorstCase, we find that at a 95$\%$ confidence level, the the precision of parameters derived from the second set are comparable to those obtained from the first set.
However, for MBHB sources of lower intensity, such as Q3-delay-1 and Q3-delay-4, when comparing the first set IdealCase with the second set WorstCase, we find that at a 95$\%$ confidence level, the precision of parameters derived from the second set are significantly worse than those from the first set. 

In this paper, the method of wavelet decomposition and reconstruction is employed. For lower-intensity MBHB sources, the precision of luminosity distance, represented by the ratio ($\frac{\Delta D_L}{D_L}$) at the $95\%$ confidence level, is enhanced when utilizing wavelet decomposition and reconstruction methods for the analysis of overlapping gravitational wave (GW) signals from MBHB and DWD sources, with improvements by factors of two and three, corresponding to Q3-delay-1 and Q3-delay-4, respectively. The angular resolution $ \Delta \Omega_s $ derived from the third set WorstCase(DWT) is enhanced by $\sim$ 20 and $\sim$ 40 times when employing wavelet decomposition and reconstruction methods, compared to those from the second set WorstCase, corresponding to Q3-delay-1 and Q3-delay-4, respectively. The wavelet decomposition and reconstruction method can significantly mitigate the impact of overlap with DWDs for lower-intensity MBHB sources.   

The angular resolution and the precision of luminosity distance of the method based on a short observation time are higher than those in \cite{Ruan:2019tje}, which is based on an observation time of about 10 days. Through the distance-redshift relation, the luminosity distance uncertainty obtained from gravitational wave sources can help constrain the accuracy of the Hubble constant in cosmology \cite{Song:2022siz}. The precision of luminosity distance, represented by the ratio ($ \lg \left( \Delta D_L / D_L \right) $) at the 95\% confidence level for Q3-delay-3 and Q3-delay-5 for Taiji is higher than that for ET2CE \cite{Song:2022siz}, which is about $ \lg \left( \Delta D_L / D_L \right)\approx -2 $. The precision of luminosity distance, represented by the ratio ($ \lg \left( \Delta D_L / D_L \right) $) at the 95\% confidence level for Q3-delay-2 for Taiji is higher than that for ET \cite{Song:2022siz}, which is about $ \lg \left( \Delta D_L / D_L \right)\approx -1.5 $. The luminosity distance uncertainty at the 95$\%$ confidence level obtained by MBHB, based on the method presented in this paper, can enhance the constraints on the accuracy of the Hubble constant in cosmology.
\section*{Acknowledgements}
We thank the anonymous referees for very helpful suggestions, which are used to improve the manuscript.
This work has been supported by the Fundamental Research Funds for the Central Universities.
This work has been supported in part by the National Key Research and Development Program of China under Grant No.2020YFC2201500, the National Science Foundation of China (NSFC) under Grants Nos. 12147103, 11821505, 11975236, and 12235008, the Strategic Priority Research Program of the Chinese Academy of Sciences under Grant No. XDB23030100. In this paper, The numerical computation of this work was completed at TAIJI Cluster of University of Chinese Academy of Sciences. 
\bibliography{ref,library}
\bibliographystyle{apsrev}
\end{document}